# Tangled magnetic field model of QPOs

Joseph Bretz,[1]* C. A. van Eysden,[1,2]† Bennett Link[1]‡

[1]*Department of Physics, Montana State University, Bozeman, MT 59717, USA*
[2]*School of Natural Sciences, University of Tasmania, Cradle Coast Campus, Burnie, TAS 7320, Australia*

## ABSTRACT

The highly tangled magnetic field of a magnetar supports shear waves similar to Alfvén waves in an ordered magnetic field. Here we explore if torsional modes excited in the stellar interior and magnetosphere can explain the quasi-periodic oscillations (QPOs) observed in the tail of the giant flare of SGR 1900+14. We solve the initial value problem for a tangled magnetic field that couples interior shear waves to relativistic Alfvén shear waves in the magnetosphere. Assuming stellar oscillations arise from the sudden release of magnetic energy, we obtain constraints on the energetics and geometry of the process. If the flare energy is deposited initially inside the star, the wave energy propagates relatively slowly to the magnetosphere which is at odds with the observed rise time of the radiative event of $\lesssim$ 10 ms. Nor can the flare energy be deposited entirely outside the star, as most of the energy reflects off the stellar surface, giving surface oscillations of insufficient magnitude to produce detectable modulations of magnetospheric currents. Energy deposition in a volume that straddles the stellar surface gives agreement with the observed rise time and excites a range of modes with substantial amplitude at observed QPO frequencies. In general, localized energy deposition excites a broad range of modes that encompasses the observed QPOs, though many more modes are excited than the number of observed QPOs. If the flare energy is deposited axisymmetrically, as is possible for a certain class of MHD instabilities, the number of modes that is excited is considerably reduced.

## 1 INTRODUCTION

Soft Gamma-ray Repeaters (SGRs) are magnetars that flare in the X-ray and soft gamma-ray bands. The three observed SGR giant flares released $10^{43}$ to $10^{47}$ ergs in radiation[1]; each flare began with a sharp (unresolved) luminosity peak followed by a decrease in luminosity into the tail of the flare which lasts hundreds of seconds (Barat et al. 1979; Hurley et al. 1999, 2005).

The first giant flare was produced by SGR 0526–66 on 1979 March 5 (Barat et al. 1979; Mazets et al. 1979; Cline et al. 1980), releasing ∼ 2×10^{45} ergs (Fenimore et al. 1996). The 1998 August 27 giant flare of SGR 1900+14 released ≳ 4×10^{43} ergs with a rise time of ≲ 4 ms to a peak of width ∼ 1 s (Feroci et al. 1999; Hurley et al. 1999). The 2004 December 27 giant flare of SGR 1806–20 released ≳ 4×10^{46} ergs (Hurley et al. 2005; Palmer et al. 2005; Terasawa et al. 2005). Measured spin-down parameters imply a strength of the magnetic dipolar component at the stellar surface of 6×10^{14} G for SGR 0526–66 (Tiengo et al. 2009), 7 × 10^{14} G for SGR 1900+14 (Mereghetti et al. 2006), and 2×10^{15} G for SGR 1806–20 (Nakagawa et al. 2008).

The energetics of giant flares indicate that these events are driven by the enormous magnetic energy reservoir of over $10^{48}$ ergs possessed by a magnetar (Thompson & Duncan 1995, 2001; Thompson et al. 2002). In the fireball model of magnetar flares (Thompson & Duncan 1995, 2001), a giant flare arises when an evolving helical twist of the core magnetic field stresses the crust until it suddenly fails. The energy in the twist then propagates as Alfvén energy into the magnetosphere, where shearing and reconnection of the magnetic field produces a fireball of electron-positron pairs confined by the magnetic field near the stellar surface. Decay of the fireball produces the flare tail for ∼ 100 s. For SGR 1900+14, the width of the initial peak of ∼ 1 s is attributed to the Alfvén crossing time of the core.

The initial rise over ≲ 4 ms is not accounted for quantitatively in the fireball model of Thompson & Duncan (1995, 2001). Lyutikov (2006) argued that the quick rise requires the energy that drives the flare to be released directly into the magnetosphere. Lyutikov proposed that the evolving magnetic field of the core and crust slowly twists the magnetosphere until it becomes dynamically unstable. A tearing mode instability develops over a time scale of $R_*/c \sim 30\,\mu s$, fast enough to explain the initial rise in luminosity over ∼ 4 ms. In support of Lyutikov's conclusions, Link (2014) showed that the jump in wave speed between the stellar interior and the magnetosphere creates an impedance mismatch that confines the energy to the stellar core for a second or longer (depending on the magnetic field strength).

Another important aspect of giant flares is the presence of quasi-periodic oscillations (QPOs). The giant flares in both SGR 1900+14 and SGR 1806−20 showed QPOs at frequencies in the $\sim 10$–$100$ Hz range with a spacing of $\sim 10$ Hz. SGR 1900+14's QPOs were found to be $28 \pm 2$ Hz, $53 \pm 5$ Hz, 84 Hz (width unmeasured), and $155 \pm 6$ Hz (Strohmayer & Watts 2005). The QPOs of SGR 1806−20 were detected at $18 \pm 2$ Hz, $26 \pm 3$ Hz, $30 \pm 4$ Hz, $93 \pm 2$ Hz, $150 \pm 17$ Hz, $626 \pm 2$ Hz, and $1837 \pm 5$ Hz (Israel et al. 2005; Strohmayer & Watts 2006; Watts & Strohmayer 2006; Hambaryan et al. 2011). A recent reanalysis of the SGR 1806−20 giant flare by Miller et al. (2019) found additional QPOs at $51 \pm 0.52$ Hz, $97 \pm 5.94$ Hz, and $157 \pm 1.14$ Hz. These QPOs are naturally interpreted as normal modes of the neutron star-magnetosphere system, with modulation arising as movement of magnetic footpoints on the stellar surface drives magnetospheric currents that create variations in the gamma-ray optical depth (Timokhin et al. 2008); beaming can amplify this effect (D'Angelo & Watts 2012). Pumpe et al. (2018) found evidence for oscillations in SGR 1900+14 at 7.7 Hz.

* joseph.bretz@montana.edu
† anthony.vaneysden@utas.edu.au
‡ link@montana.edu
[1] All energy estimates assume isotropic emission.

Many models have been advanced to explain QPOs as originating from stellar oscillations. Most work has assumed smooth magnetic field geometries, such as that of a dipolar field or some variant (Glampedakis et al. 2006; Levin 2006, 2007; Sotani et al. 2008b,a; Cerdá-Durán et al. 2009; Colaiuda et al. 2009; Colaiuda & Kokkotas 2011, 2012; Gabler et al. 2011, 2012, 2013b,a, 2014; van Hoven & Levin 2011, 2012; Passamonti & Lander 2013; de Souza & Chirenti 2018). These smooth field geometries lead inevitably to the existence of an Alfvén continuum in the stellar core that resonantly absorbs elastic energy from the crust, damping the crust motion in under 0.1 seconds. van Hoven & Levin (2011) have shown that the presence of gaps in the Alfvén continuum prevents the resonant absorption of certain modes.[2]

Dipolar and other smooth field configurations as considered in the aforementioned work are dynamically unstable; linked poloidal and toroidal components are needed to stabilized the star, as in the nearly axisymmetric "twisted torus" (Braithwaite & Nordlund 2006). Simulations of turbulent initial conditions indicate the magnetic field generally evolves into a stable, tangled equilibrium (Braithwaite 2008), topologically distinct from a smooth field configuration. van Hoven & Levin (2011) and Link & van Eysden (2016) (hereafter LvE) found that a magnetic tangle breaks up the Alfvén continuum into discrete modes, and they estimated the frequencies for an isotropically tangled magnetic field. By matching a simple model of a tangle plus a dipole to observations, LvE showed that the magnetic tangle sufficiently dominates the dipole field in SGR 1900+14 (hereafter SGR 1900) to eliminate the problematic Alfvén continuum that arises when a smooth field configuration is assumed. By contrast, LvE estimate that the tangled component of SGR 1806−20 is comparable in strength to that of the dipole field. Tangled magnetic fields have been considered elsewhere in studies of QPOs in magnetars (Sotani 2015) and X-ray emission from central compact objects (Gourgouliatos et al. 2020).

In this paper, we apply the tangled field model of LvE to SGR 1900's giant flare. Under the interpretation that the observed QPOs represent stellar oscillations, we constrain where in the star the magnetic energy that ultimately produces the flare is deposited and how much energy is required. We consider plausible excitation geometries and compare Fourier power spectra of excited oscillations with the observed QPOs from SGR 1900's giant flare, arriving at constraints on the flare energetics.

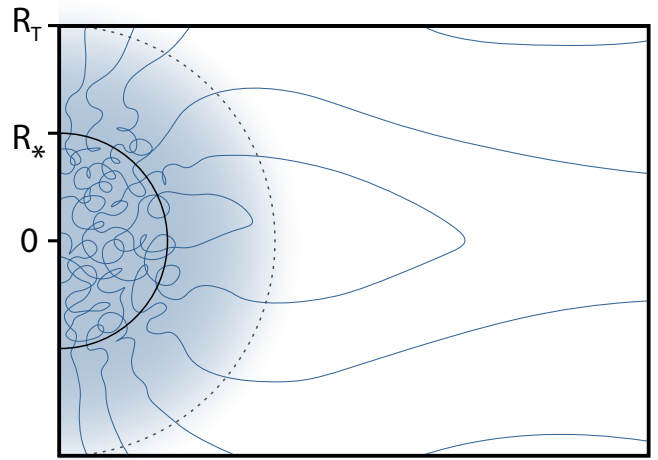

**Figure 1.** A sketch of the transition between the interior magnetic tangle and dipolar magnetosphere. Inside the star ($r < R_*$) the field is uniformly tangled. $R_T \sim 2R_*$ demarcates the transition zone where high-order multipoles die off, giving way to a dipolar field. The magnetosphere in our model supports isotropic stress at all $r > R_*$; our model does not include a transition to a smooth magnetic geometry.

We first present the model and its limitations in §2. The model's dynamics are dominated by the impedance mismatch at the stellar surface which is addressed in §2.1. In §2.2 we solve the initial value problem for initial conditions defined in §2.3. The detectability criterion for excited modes is discussed in §3, and Fourier power is calculated using the method presented in §4. Constraints resulting from our model are laid out in §5. We discuss our conclusions in §6.

## 2 COUPLED TANGLE MODEL OF QPOs

*The stellar interior.* LvE estimate that for SGR 1900, higher-order magnetic multipoles have about 14 times as much energy as the dipole component. As a simplification, we ignore the dipole component, and treat the magnetic field inside the star as uniformly and isotropically tangled. The magnetic tangle gives an isotropic contribution to the shear stress with an effective shear modulus (for normal protons) $\mu_B \equiv \langle B_t^2 \rangle / (4\pi)$, where $\langle B_t^2 \rangle^{1/2}$ is the spatially averaged tangled magnetic field (LvE; see also Hosking et al. 2020) for a detailed treatment of tangled fields in MHD).[3] For SGR 1900, LvE estimate $\mu_B \approx 6 \times 10^{29}$ ergs cm$^{-3}$, while the volume-averaged shear modulus of the relatively thin solid crust is about half this value. LvE showed that crust rigidity has a negligible effect on stellar shear oscillations; rather, the oscillation frequencies are better treated as mass-loaded magnetic oscillations. Hence, we ignore crust rigidity. We also ignore the effects of realistic stellar structure, treating the star as having constant mass density, $\rho = 3M_*/4\pi R_*^3$, where $M_*$ is the stellar mass and $R_*$ the radius. The wave propagation speed in the stellar interior is $c_A \sim 10^8$ cm s$^{-1}$ (see below). The thickness of crust is $\Delta R \sim 0.05R$.

---

[2] If one of the crust's natural frequencies lies within a continuum gap, then the coupled magnetic field does not resonantly absorb that mode. In addition "edge modes" at the edges of the continua appear and do not get absorbed as quickly.

[3] If the core protons are superfluid, the effective shear modulus becomes $H_{c1}B/4\pi$, where $H_{c1} \simeq 10^{15}$ G is the lower critical field. Since $H_{c1}$ is comparable to field magnitudes of interest, the distinction between normal and superfluid protons in the core is not an important one for the normal mode problem.

For wavelengths longer than the crust thickness, corresponding to frequencies below $c_A/\Delta R \sim 2$ kHz (which encompasses all low-frequency QPOs), realistic crust structure will have a minor effect on wave propagation (see also, Link 2014); we therefore treat the stellar surface as a sharp discontinuity in the density. We will consider geometries for the energy deposition in which the radial dimension is always larger than the crust thickness. LvE concluded that the magnetic dipole moment cannot be ignored in SGR 1806, so the isotropic tangle model we develop in this paper cannot be applied to that object. SGR 1900, by contrast, has a weaker dipolar component of the magnetic field, and so the magnetic tangle is the primary feature of the mode problem.

LvE determined the tangle strength of SGR 1900 by ascribing the lowest, zero-angular-momentum mode with the 28 Hz, SGR-1900 QPO. We do the same here, however coupling to a magnetosphere shifts the frequencies of the modes. Aligning the lowest mode of the coupled system with 28 Hz gives a slightly weaker interior magnetic tangle strength of $2.3 \times 10^{15}$ G which implies the tangle has about 11 times more magnetic energy than the dipole component.

The magnetic tangle supports shear waves which we will refer to as "Alfvén waves," though usually this term refers to shear waves that propagate along the magnetic field; in the approximation of a magnetic tangle, there is no preferred direction. The neutron fluid is expected to be in the superfluid state, with negligible coupling to the motion of the protons and electrons. The propagation speed for Alfvén waves inside the star is then $c_A = \sqrt{\mu_B/x_p\rho}$, where $x_p \approx 0.1$ is the mass fraction of protons. The time scale for Alfvén waves to cross the star is

$$t_A = \frac{2R_*}{c_A} = 0.025 \text{ s } \left(\frac{x_p}{0.1}\right)^{1/2} \left(\frac{R_*}{10 \text{ km}}\right)^{-1/2}$$
$$\times \left(\frac{M_*}{1.4 M_\odot}\right)^{1/2} \left(\frac{\langle B_t^2 \rangle^{1/2}}{2.3 \times 10^{15} \text{G}}\right)^{-1}.$$

The rotation period of a magnetar is much longer than the crossing time scale, so we ignore rotation and treat the star as a sphere of radius $R_*$. We do not expect our model to apply for QPOs at higher frequencies where modes with shorter wavelengths can probe the structure of the magnetic tangle, see LvE.

*Magnetosphere.* In the magnetosphere, shear disturbances propagate as relativistic Alfvén waves. The field remains tangled near the stellar surface and becomes mostly dipolar at $\sim 2R_*$ (see Fig. 1). As the field becomes smooth, Alfvén waves can propagate only along the magnetic field. Alfvén wave energy can escape to infinity along open field lines and will be returned to the star along closed field lines. Since the structure of the magnetosphere is not well known, as a simplification we treat the magnetosphere as having isotropic magnetic stress, so that shear waves propagate at the speed of light in any direction. In the long-wavelength description in which we will work, we set the effective shear modulus of the spherically symmetric magnetosphere equal to the value determined by the observationally inferred *dipole* field of strength $7 \times 10^{14}$ G. Hence, at the stellar surface there is a jump in both the wave propagation speed and the effective shear modulus. The multi-pole contributions to the magnetic tangle decay over a length scale of $\sim R_*$ in the magnetosphere. For a typical QPO of frequency 100 Hz, the wavelength in the magnetosphere is $300R_*$, so we treat the transition from the magnetic tangle to a smooth, dipole-like configuration as a discontinuous jump at the stellar surface from the field strength of the tangle to the dipole value. In this crude treatment, the magnetosphere has a constant magnetic field strength everywhere, and no outer boundary. Inside the star, typical wavelengths of the modes we are interested in are of order the stellar radius or smaller, so this long wavelength description is a rather crude approximation. In a real magnetosphere, the magnetic stresses will be anisotropic, which we expect will cause coupling of Alfvén waves to outside the star to be somewhat less efficient than we implicitly assume by treating the magnetic stresses in the magnetosphere as isotropic. As we show below, wave energy reaches the magnetosphere from the stellar interior with very low efficiency. The main cause for this impedance mismatch is the jump in Alfvén wave propagation speed across the stellar surface by about two orders of magnitude, due to the jump in density.

We estimate the effects of general relativity by introducing a red-shift factor, $z_r \equiv (1 - R_s(M_*)/R_*)^{1/2}$, where $R_s$ is the Schwarzschild radius, when evaluating oscillation frequencies at infinity; $z_r$ is about 0.77 for fiducial stellar parameters.

The stellar interior supports shear waves with frequencies beginning at $\sim 20$ Hz (LvE). Gravity modes (g modes), involve radial motion with buoyancy as the restoring force, and have much higher frequencies, about 500 Hz (Kantor & Gusakov 2014). Since g modes have much higher frequencies than the low-frequency QPOs, we neglect them. Sound waves in the core propagate at nearly the speed of light (e.g., Bedaque & Steiner (2015)), and have frequencies beginning above $\sim 10$ kHz, so we neglect them as well. We therefore consider only shear waves in the stellar core. We neglect possible coupling of shear waves to magnetosonic waves in the magnetosphere, and consider only relativistic Alfvén waves in the magnetosphere.

The star plus magnetosphere has discrete normal modes that slowly decay as Alfvén energy propagates to infinity; as we show, the time scale for the loss of energy to infinity is much longer than the flare duration of $\sim 100$ s. Another dissipative mechanism is electron viscosity, for which we estimate the damping timescale, $\tau_{ev}$, for a density of $4 \times 10^{14}$ g/cm³ at $10^9$ K. The electron viscosity is $\eta = 10^{17}$ g cm⁻¹ s⁻¹ (Flowers & Itoh 1979). Damping by shear velocity is very slow:

$$\tau_{ev} = \frac{\rho}{\eta} \frac{1}{k^2} \approx 3 \text{ years} \left(\frac{\lambda}{R_*}\right)^2 \left(\frac{T}{10^9 \text{K}}\right)^2.$$

where $\lambda = 2\pi/k$ is the wavelength. We ignore electron viscosity henceforth. Other damping processes we have not considered, such as dissipation of induced electrical currents, might play a more significant role than electron viscosity.

### 2.1 Impedance Mismatch Between the Stellar Interior and the Magnetosphere

Before addressing the general initial-value problem, we demonstrate that there is a very strong impedance mismatch between the stellar interior and the magnetosphere that strongly inhibits the flow of Alfvén wave energy through the stellar surface. This mismatch was examined by Link (2014) for a planar geometry; here we show that the mismatch is even greater in spherical geometry.

We are interested in modes with frequencies in the range of 10-100 Hz; only shear waves have frequencies in this range. For shear motions, mass motion is confined primarily to gravitational equipotentials, so the radial motion is zero. The displacement vector $\boldsymbol{u}(\boldsymbol{r}, t)$ of a mass element in either the stellar interior or the magnetosphere obeys

$$\nabla \cdot \boldsymbol{u}(\boldsymbol{r}, t) = 0, \tag{1}$$

and the wave equation

$$c_A^2 \nabla^2 \boldsymbol{u} - \partial_t^2 \boldsymbol{u} = 0, \tag{2}$$

where $c_A$ is the wave speed. In spherical coordinates $(r, \theta, \phi)$, a normal mode of frequency $\omega$ takes the form

$$\boldsymbol{u}(\boldsymbol{r}, t) = \left[ u_\theta(\boldsymbol{r}, \omega)\hat{\theta} + u_\phi(\boldsymbol{r}, \omega)\hat{\phi} \right] e^{-i\omega t}. \tag{3}$$

Assuming a separable solution, eqs.(1) and (2) give

$$u_\phi(\boldsymbol{r}, t = 0) = w(r, \omega) \frac{\partial}{\partial \theta} Y_{\ell m}(\theta, \phi) \qquad u_\theta(\boldsymbol{r}, t = 0) = -w(r, \omega) \frac{1}{\sin \theta} \frac{\partial}{\partial \phi} Y_{\ell m}(\theta, \phi), \tag{4}$$

where the $Y_{\ell m}$ are spherical harmonic functions. The radial function $w(r, \omega)$ satisfies

$$\left( \frac{d^2}{dr^2} + \frac{2}{r} \frac{d}{dr} - \frac{\ell(\ell + 1)}{r^2} + k^2 \right) w(r, \omega) = 0, \tag{5}$$

where $k \equiv \omega / c_A$. The solutions are spherical Bessel and Neumann functions, $j_\ell(kr)$ and $n_\ell(kr)$ and their linear combinations, the spherical Hankel functions $h_\ell^{(1)}(kr) = j_\ell(kr) + i\, n_\ell(kr)$ and $h_\ell^{(2)}(kr) = j_\ell(kr) - i\, n_\ell(kr)$.

We denote the region inside the star by the subscript $<$, and the region outside the star by the subscript $>$. At the stellar surface ($r = R_*$), the displacement and the traction must be continuous, giving the boundary conditions,

$$w_<(R_*) = w_>(R_*) \tag{6}$$

$$\mu_< \left( \frac{dw_<}{dr} - \frac{w_<}{r} \right)_{r=R_*} = \mu_> \left( \frac{dw_>}{dr} - \frac{w_>}{r} \right)_{r=R_*}. \tag{7}$$

We take the $>$ region to be infinite. The interior solution that is bounded at $r = 0$ is $j_\ell(kr)$. To evaluate the rate of energy flow through the stellar surface to infinity, we assume a solution for $r > R_*$ that becomes an outgoing wave for large $r$, that is, $h_\ell^{(1)}(kr)$. The solution is

$$w_\ell^<(r) = j_\ell(k_< r) \qquad w_\ell^>(r) = A_\ell\, h_\ell^{(1)}(k_> r). \tag{8}$$

The coefficients $A_\ell$ are set by the boundary conditions. Because this solution gives an energy flow to infinity, $\omega$ is complex and discrete.

Define $z \equiv k_< r$, $z_* \equiv k_< R_*$, $\alpha \equiv \mu_> / \mu_<$, and $\gamma \equiv c_< / c$. Combining eqs.(6) and (7) gives

$$\left( \frac{d \ln j_\ell(z)}{dz} - \frac{1}{z} \right)_{z_*} = \alpha \left( \frac{d \ln h_\ell^{(1)}(\gamma z)}{dz} - \frac{1}{z} \right)_{z_*}. \tag{9}$$

This equation has an infinite sequence of solutions which we denote $z_{\ell n}$, where $n = 0, 1, 2, 3, \ldots$ is the overtone number and corresponds with the number of radial nodes inside the star.

Numerical solutions to eq.(9) show the two classes of modes plotted in Fig. 2. The modes split into two distinct types: the type-I modes (left panel) that have very low decay rates, and the type-II modes (right panel) that have higher frequencies and high damping rates. The type-II modes are generally overdamped since their eigenfunctions have significant amplitude for $r > R_*$. Consequently, they are strongly excited when energy is deposited outside the star, and externally-deposited energy can be lost to infinity at the speed of light, damping at a rate of $\sim c/R_*$. Due to their kilohertz frequencies, type-II modes are not candidates for QPOs. The typical giant flare tail duration of 300 seconds, demarcated by the horizontal gray line in Fig. 2, is much longer than the decay rates of the type-I modes. The type-II modes decay within milliseconds, and so are not dynamically relevant for most of the flare tail duration; we henceforth focus on the type-I modes.

The wave speed inside the star is small compared to $c$:

$$\gamma \equiv \frac{c_<}{c} = \frac{1}{c} \sqrt{\frac{\langle B_t^2 \rangle}{4\pi \rho_d}} \approx 0.0027 \left( \frac{R_*}{10\ \text{km}} \right)^{3/2} \left( \frac{M_*}{1.4 M_\odot} \right)^{-1/2} \left( \frac{x_p}{0.1} \right)^{-1/2} \left( \frac{\langle B_t^2 \rangle^{1/2}}{2.3 \times 10^{15}\ \text{G}} \right), \tag{10}$$

where $\rho_d \equiv x_p \rho$ is the total proton density, $x_p$ is the fraction of protons, and $\rho$ is the total mass density. We assume the neutrons are superfluid, so that only the proton density $x_p \rho$ determines the wave speed, independent on whether or not the protons are superconducting or normal (Chamel & Haensel 2006). With the idealization that the field goes suddenly from a strong interior tangle ($B_t$) to a weaker tangled magnetosphere (of strength equal to SGR 1900's dipole field, $B_d$) at the stellar surface, the quantity $\alpha \equiv \mu_> / \mu_<$ is also small:

$$\alpha \equiv \frac{\mu_>}{\mu_<} = \frac{B_d^2}{\langle B_t^2 \rangle} \approx 0.089 \left( \frac{B_d}{7 \times 10^{14}\ \text{G}} \right)^2 \left( \frac{\langle B_t^2 \rangle^{1/2}}{2.3 \times 10^{15}\ \text{G}} \right)^{-2}. \tag{11}$$

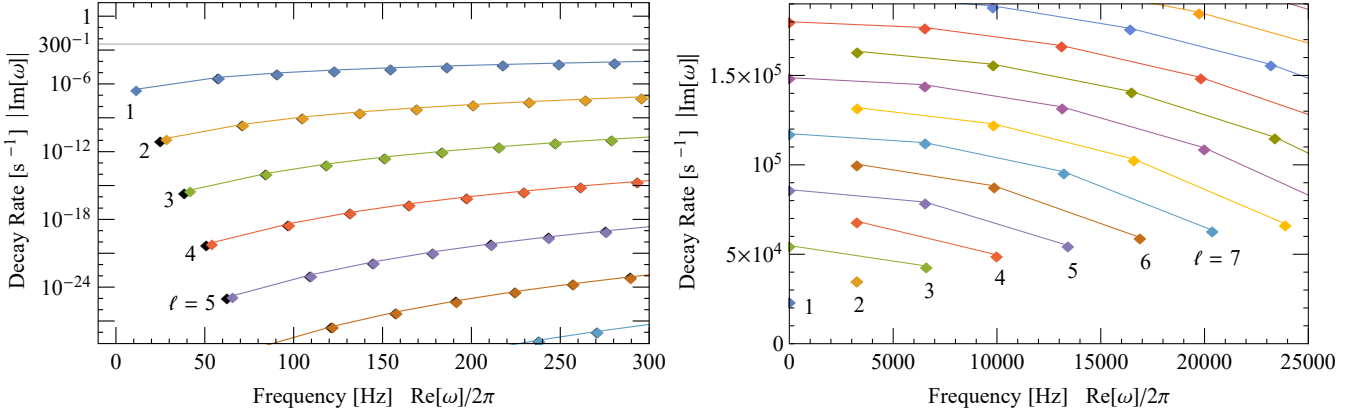

**Figure 2.** Mode decay rate versus frequency. The numerical solutions to eq.(9) are colored and connected by $\ell$. The redshift factor of $z_r$ is included. **Left:** The type-I modes consist of a fundamental mode of 10's of Hz for each $\ell$ followed by a chain of overtones with higher frequency and decay rate. The approximate eigenfrequencies, eq.(14), are black and underlay the numerical solutions. The deviation between the approximate and numerical solutions appears greatest for the fundamental for each $\ell$ (the $\ell = 1$ fundamental approximation is omitted, because this mode has non-zero angular momentum). The horizontal gray line marks the typical giant flare tail duration. **Right:** The type-II modes have decay rates $> 10^3$ s$^{-1}$. The number of eigenfrequencies per $\ell$ is $\lceil \ell/2 \rceil$. The odd $\ell$ modes have an eigenfrequency with vanishing real part. These imaginary eigenfrequencies correspond with overdamped modes.

We obtain approximate analytic solutions for the dynamically relevant, type-I modes. The type-I modes have small imaginary components, $\mathrm{Im}(z) \ll 1/\gamma$, so we expand the right side of eq.(9) to leading order in the small parameters $\alpha$ and $\gamma$. The zeroth order solution in $\alpha$ is given by

$$\left( \frac{d \ln j_\ell(z)}{dz} - \frac{1}{z} \right)_{z_*} = 0 \, .$$

We denote the solutions by $z_{\ell n}^0$. For small $\alpha$ and $\gamma$, the solution acquires a small imaginary part. Writing the solutions for non-zero $\alpha$ as $z_{\ell n}^0 + \delta z$, expanding the left side of eq.(9) to first order in $\delta z$ gives

$$\left( \frac{d \ln j_\ell(z)}{dz} - \frac{1}{z} \right)_{z_{\ell n}^0 + \delta z} \approx -\delta z \left[ 1 - \frac{(\ell+2)(\ell-1)}{(z_{\ell n}^0)^2} \right] \, .$$

Expanding the right side of eq.(9) to leading order in $\alpha$ and $\gamma$ and combining with eq.(13) eventually gives:

$$\omega_{\ell n} = \frac{c_<}{R_*} z_r z_{\ell n}^0 \left( 1 - i \, \alpha \, \gamma^{2\ell+1} \frac{(z_{\ell n}^0)^{2\ell-1}}{[(2\ell-1)!!]^2} \left[ 1 - \frac{(\ell+2)(\ell-1)}{(z_{\ell n}^0)^2} \right]^{-1} \right) .$$

We have included the redshift factor $z_r$ for comparison with later results. Note the very strong scaling $\mathrm{Im}[\omega_{\ell n}] \sim \gamma^{2\ell+1}$. The solutions in eq.(14) are approximations of the type-I modes and are plotted in black under the numerical solutions in Fig. 2. The approximate fundamental frequencies for each $\ell$ deviate the most from the numerical solutions in the plotted range.

The energy $E_{\ell n}$ in a type-I mode is lost to infinity at a rate

$$\frac{\dot{E}_{\ell n}}{E_{\ell n}} \sim \mathrm{Im}(\omega_{\ell n}) \approx -\frac{c_<}{R_*} z_r \alpha \, \gamma^{2\ell+1} \frac{(z_{\ell n}^0)^{2\ell}}{[(2\ell-1)!!]^2} \, .$$

The $\ell = 1$ modes decay the quickest of the type-I modes, however they do not receive any power for initial conditions with zero net external torque and zero angular momentum, as we show below. Thus the $\ell = 2$, type-I modes lose energy to infinity the quickest. The $\ell = 2$ fundamental mode $\omega_{20}$ has a lifetime of $\sim 2000$ years. The higher-frequency overtones decay more quickly with the fifth overtone $\omega_{25}$ decaying over $\sim 2$ years. All type-I modes persist for much longer than the typical giant flare duration of 300 seconds.

The very low rate of energy flow to infinity is due to the discontinuity in propagation speed at the stellar surface, and the properties of the eigenfunctions in the spherical geometry which die off very quickly with $r$. Near the stellar surface, the solution outside the star takes the form:

$$w_\ell^>(k_> r) = A_\ell h_\ell^{(1)}(k_> r) = \frac{j_\ell(k_< R_*)}{h_\ell^{(1)}(k_> R_*)} h_\ell^{(1)}(k_> r) \longrightarrow j_\ell(k_< R_*) \left( \frac{r}{R_*} \right)^{-(\ell+1)} .$$

For large $r$ such that $k_> r \gg 1$, corresponding to $r \gg k_<^{-1} \sim \gamma^{-1} R_* \sim 10^2 R_*$, the solution goes as $h_\ell^{(1)} \to \gamma^{\ell+1}/r$, so the amplitude is low.[4] The energy flux scales as $r^2 k_> [h_\ell^{(1)}(k_> r)]^2 / h_\ell^{(1)}(k_> R_*) \propto k_>^{2\ell+1} \propto \gamma^{2\ell+1}$, as given by the imaginary part of the frequency.

While this solution gives a wave amplitude outside the star that is large only near the stellar surface, a solution that accounts for the transition from a magnetic tangle to a dipole at large $r$ could be significantly different.

---

[4] At very large $r$, the solution grows exponentially, a consequence of assuming a radiative normal mode solution.

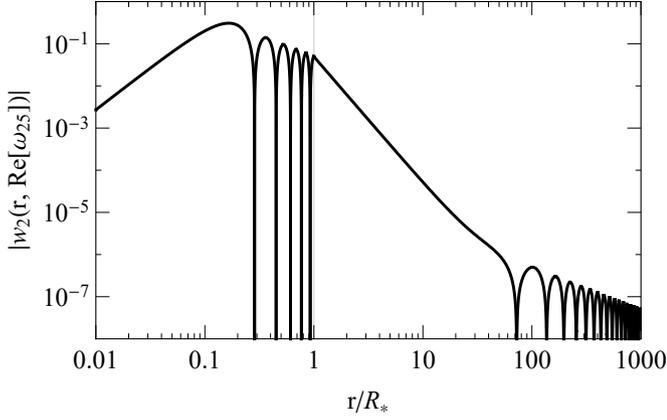

**Figure 3.** Plot of the radial dependence of an eigenfunction with $\omega = Re\,[\omega_{25}]$, see eq.(20). The vertical gray line denotes the stellar surface.

## 2.2 Initial-value Formulation

To solve for the time-dependent dynamics of the star and magnetosphere, we construct a solution as a continuous superposition of normal modes

$$\boldsymbol{u}\,(\boldsymbol{r},t) = \int_0^\infty a(\omega)\bar{\boldsymbol{u}}\,(\boldsymbol{r},\omega)\cos(\omega t)\,d\omega \qquad t \geq 0. \tag{17}$$

where $\omega$ is real, and we have assumed, with minor loss of generality, that the matter has no kinetic energy at $t = 0$; all of the energy is elastic. The coefficients $a(\omega)$ are determined by the initial conditions. Each Fourier component $\bar{\boldsymbol{u}}$ obeys the wave equation

$$c_A^2\nabla^2\bar{\boldsymbol{u}} + \omega^2\bar{\boldsymbol{u}} = 0\,, \tag{18}$$

where the propagation speeds, $c_A$, are the same as before (see eq.(10)), $c_< = \gamma c$ inside the star, and $c_> = c$ outside. We will consider boundary conditions that allow separation of the angular coordinates:

$$\bar{u}_\phi(\boldsymbol{r},\omega) = w(r,\omega)\frac{\partial}{\partial\theta}Y_{\ell m}(\theta,\phi) \qquad \bar{u}_\theta(\boldsymbol{r},\omega) = -w(r,\omega)\frac{1}{\sin\theta}\frac{\partial}{\partial\phi}Y_{\ell m}(\theta,\phi)\,, \tag{19}$$

and $\bar{u}_r = 0$.

As basis eigenfunctions, we take superpositions of outgoing and incoming waves,

$$w_\ell^<(r,\omega) = j_\ell\left(\frac{r\omega}{c_<}\right) \qquad w_\ell^>(r,\omega) = \frac{1}{2}\left[A_\ell(\omega)\,h_\ell^{(1)}\left(\frac{r\omega}{c_>}\right) + B_\ell(\omega)\,h_\ell^{(2)}\left(\frac{r\omega}{c_>}\right)\right] \tag{20}$$

The coefficient $B_\ell(\omega)$ is related to $A_\ell(\omega)$ by the boundary conditions eqs.(6) and (7). As shown in Appendix C, for real $\omega$,

$$B_\ell(\omega) = A_\ell^*(\omega)\,. \tag{21}$$

$A_\ell$ and $B_\ell$ have the same modulus, and the solution is a standing wave, with the exterior solution given by

$$w_\ell^>(r) = \mathrm{Re}\left[A_\ell(\omega)h_\ell^{(1)}\left(\frac{r\omega}{c_>}\right)\right]. \tag{22}$$

An example eigenfunction is plotted in Fig. 3.

We write eq.(17) in terms of the eigenfunctions:

$$\boldsymbol{u}_i\,(\boldsymbol{r},t) = \sum_\ell\sum_{m=-\ell}^{\ell}\int_0^\infty a_{\ell m}(\omega)\bar{\boldsymbol{u}}_{i\ell m}\,(\boldsymbol{r},\omega)\cos(\omega t)\,d\omega\,, \tag{23}$$

where $\bar{\boldsymbol{u}}_{\ell m}$ are the eigenfunctions given by eqs.(19) and (20).

Using the orthogonality condition eq.(C17) derived in Appendix C, the Fourier coefficients are

$$\begin{aligned}a_{\ell m}(\omega) &= N_\ell(\omega)\left[\rho_<\int_0^{R_*} f_r(r)w_\ell^<\left(\frac{\omega r}{c_<}\right)r^2dr + \rho_>\int_{R_*}^\infty f_r(r)w_\ell^>\left(\frac{\omega r}{c_>}\right)r^2dr\right]\\ &\quad\times \int_0^{2\pi}\int_0^\pi\left[\frac{\partial^2 f_\theta}{\partial\theta^2}\frac{\partial f_\phi}{\partial\phi}\frac{\partial Y_{\ell m}^*}{\partial\theta} + \frac{1}{\sin^2\theta}\frac{\partial f_\theta}{\partial\theta}\frac{\partial^2 f_\phi}{\partial\phi^2}\frac{\partial Y_{\ell m}^*}{\partial\phi}\right]\sin\theta\,d\theta\,d\phi\,,\end{aligned} \tag{24}$$

where the initial conditions $(f_r, f_\theta, f_\phi)$ are defined in §2.3. The effective mass density inside the star is $\rho_< = \rho_d$ from eq.(10). For relativistic Alfvén waves outside the star, $\rho_> = \mu_>/c^2$ where $\mu_> = B_d^2/4\pi$. The normalization is

$$N_\ell(\omega) \equiv \left[\ell(\ell+1)\frac{\pi c_>\mu_>}{2\omega^2}A_\ell(\omega)B_\ell(\omega)\right]^{-1}. \tag{25}$$

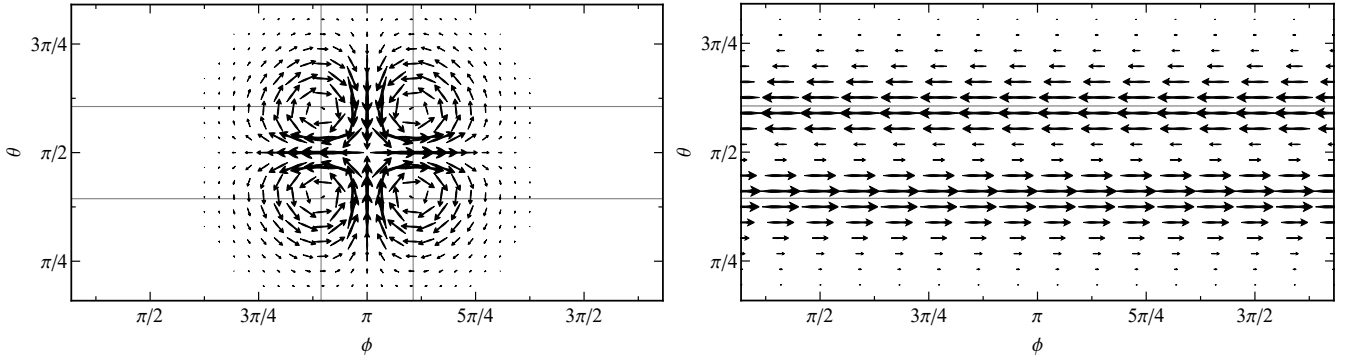

**Figure 4.** Vector plot of the angular dependence of the non-axisymmetric initial condition, eqs.(27), on the left, and the axisymmetric initial condition, eqs.(28), on the right. The characteristic extent of the initial deformation field is marked by the horizontal gray lines for $\theta_0 \pm \sigma_\theta$ and the vertical gray lines for $\phi_0 \pm \sigma_\phi$. In this figure, $\sigma_\theta = \sigma_\phi = 1/3$.

Combining eqs.(4), (8), and (23), the full solutions for the inner ($<$) and outer ($>$) regions are

$$u_{\phi<}(\boldsymbol{r},t) = \sum_{\ell=1}^{\infty}\sum_{m=-\ell}^{\ell} \frac{\partial}{\partial \theta} Y_{\ell m}(\theta,\phi) \int_0^\infty a_{\ell m}(\omega)\, j_\ell\left(\frac{\omega r}{c_<}\right)\cos(\omega t)\, d\omega\,, \tag{26}$$

$$u_{\theta<}(\boldsymbol{r},t) = -\sum_{\ell=1}^{\infty}\sum_{m=-\ell}^{\ell} \frac{1}{\sin\theta}\frac{\partial}{\partial \phi} Y_{\ell m}(\theta,\phi) \int_0^\infty a_{\ell m}(\omega)\, j_\ell\left(\frac{\omega r}{c_<}\right)\cos(\omega t)\, d\omega\,,$$

$$u_{\phi>}(\boldsymbol{r},t) = \frac{1}{2}\sum_{\ell=1}^{\infty}\sum_{m=-\ell}^{\ell} \frac{\partial}{\partial \theta} Y_{\ell m}(\theta,\phi) \int_0^\infty a_{\ell m}(\omega)\left[A_\ell(\omega)h_\ell^{(1)}\left(\frac{\omega r}{c_>}\right)+B_\ell(\omega)h_\ell^{(2)}\left(\frac{\omega r}{c_>}\right)\right]\cos(\omega t)\, d\omega\,,$$

$$u_{\theta>}(\boldsymbol{r},t) = -\frac{1}{2}\sum_{\ell=1}^{\infty}\sum_{m=-\ell}^{\ell} \frac{1}{\sin\theta}\frac{\partial}{\partial \phi} Y_{\ell m}(\theta,\phi) \int_0^\infty a_{\ell m}(\omega)\left[A_\ell(\omega)h_\ell^{(1)}\left(\frac{\omega r}{c_>}\right)+B_\ell(\omega)h_\ell^{(2)}\left(\frac{\omega r}{c_>}\right)\right]\cos(\omega t)\, d\omega\,.$$

In the above expressions we do not set $B_\ell(\omega) = A_\ell^*(\omega)$, which holds for real $\omega$; in Appendix D we express the integrals as contour integrals in the complex plane, and solve numerically.

### 2.3 Initial Conditions

Because the neutron star is isolated, the initial deformation of the star must carry zero total angular momentum. A convenient choice of initial condition that has no angular momentum, a vanishing divergence, and is also separable in spherical coordinates is obtained in Appendix A (see eqs.A7-A9):

$$u_{\phi}^{0}(\boldsymbol{r},t=0) = f_r(r)\frac{\partial^2 f_\theta(\theta)}{\partial\theta^2}\frac{\partial f_\phi(\phi)}{\partial\phi}\,, \qquad u_{\theta}^{0}(\boldsymbol{r},t=0) = -\frac{f_r(r)}{\sin(\theta)}\frac{\partial f_\theta(\theta)}{\partial\theta}\frac{\partial^2 f_\phi(\phi)}{\partial\phi^2}\,, \tag{27}$$

where $f_x(x)$ is a Gaussian-like function centered at $x_0$ with a width of $\sigma_x$ in the respective coordinate. The initial condition is normalized to an amplitude of $a$, which is related to the initial energy by eq.(30) below. This choice can be visualized as four twists in the $\phi - \theta$ plane centered at $(r_0, \theta_0, \phi_0)$ with size $\sigma_r \times r_0\sigma_\theta \times r_0\sigma_\phi$; an example is shown in Fig. 4. Four swirls in opposing directions are needed to cancel all angular momentum. The full expressions are given as eqs.(A2-A4). The system is spherically symmetric, so for simplicity we set $\theta_0 = \pi/2$ and $\phi_0 = \pi$.

We also consider axisymmetric displacements. For an axisymmetric initial condition

$$u_{\phi}^{0}(\boldsymbol{r},t=0) = f_r(r)\frac{\partial f_\theta(\theta)}{\partial\theta}\,, \qquad u_{\theta}^{0}(\boldsymbol{r},t=0) = 0\,. \tag{28}$$

Such an initial condition can be visualized as two, equal and opposite twists wrapping around the star above and below the equator ($\theta = \pi/2$), also shown in Fig. 4. $\sigma_\theta$ still specifies the extent of the twists in the $\theta$ direction.

## 3 QPO DETECTABILITY CRITERION

Our model gives the oscillation amplitude of the stellar surface as a function of the initial energy. We do not attempt to model the process that converts mechanical energy to radiation. Instead we use the estimated detectability threshold from Timokhin et al. (2008).

Timokhin et al. have shown that motion of the magnetic footpoints at the stellar surface due to stellar oscillations drives magnetospheric currents that modify the particle number density and consequently the optical depth. They estimate the amplitude of oscillations should be $\sim 0.01 R_*$ at the stellar surface to produce the observed photon luminosity modulations of $\sim 10\%$, assuming a pair plasma multiplicity of a few or larger (Beloborodov & Thompson 2007), a simplified $\ell$-dependence, and axisymmetric oscillations.

Beloborodov (2013) estimates that the multiplicity can be as large as $\sim 100$ in regions farther from the star and near the poles. The higher multiplicity reduces the detectability threshold estimate by Timokhin et al. to $\sim 0.001 R_*$. Non-axisymmetric, localized displacements amplify the currents and the subsequent modulations of the optical depth, further reducing the detectability threshold. Beaming of the emitted $\gamma$-rays could also significantly increase the observed modulation D'Angelo & Watts (2012). As a conservative upper limit for detectability, we seek to attain an amplitude of $\sim 0.01 R_*$ in our simulations.

It is important to bear in mind that our simplified magnetosphere with isotropic magnetic stress behaves very differently than the *purely dipolar* model of Timokhin et al. They calculate the magnetospheric currents at resonant Compton-scattering surfaces, *i.e.* $r \lesssim 10 R_*$, fix the value of the current there to give the observed modulation of $\sim 10\%$, which in turn implies a surface amplitude of $\sim 0.01 R_*$. In our model, the amplitude of the isotropic magnetosphere decays with $r$ as a strong power law, see eq.(16), so the amplitude at $\sim 10 R_*$ is extremely small. Neither model completely describes the magnetosphere, so we indicate the scaling of our results with surface amplitude.

## 4 METHOD

To assess the detectability of stellar oscillations for specific initial conditions, we calculate the oscillation amplitude at the stellar surface. We calculate the amplitude of each mode using eq.(23). The frequency content of the initial condition is given as $a_{\ell m}(\omega)$, and can be numerically calculated for a given $\omega$. Since the type-I modes negligibly decay over a giant flare lifetime, the spectral distribution of excited modes has negligible dependence on time. We calculate the displacement of the stellar surface at a representative "probe" spot, where the probe location,

$$\boldsymbol{r}_p \equiv (R_*, \theta_p, \phi_p),$$ (29)

is on the stellar surface. For convenience, we choose $\theta_p$ and $\phi_p$ such that $\boldsymbol{r}_p$ is at the angle of the initial condition's maximum amplitude and where $u_\theta = 0$ for all time, so only $u_\phi$ is needed to calculate the total displacement. For the localized initial condition, eq.(27), $\theta_p = \pi/2$ and $\phi_p = \pi + \sigma_\phi$. For an axisymmetric initial condition $\theta_p$ must be numerically calculated but is approximately $\pi/2 - \sigma_\theta$.

In eq.(23) the integrand has very sharp peaks at $\mathrm{Re}(\omega_{\ell n})$, because the system radiates energy to infinity very slowly, giving $\omega_{\ell n}$ a very small imaginary part. (The sharpness of the peaks enters through the frequency dependence of $A_\ell$ and $B_\ell$ in eq.(25).) Using eq.(21), we approximate each peak in the integrand as a Lorentzian of width $2 \mathrm{Im}(\omega_{\ell n})$, so that $d\omega \approx 2 \mathrm{Im}(\omega_{\ell n})$.

The physical scale of the amplitude of oscillations is determined by the elastic energy. For initial conditions that have zero initial velocity everywhere, the initial energy is all strain energy, eq.(B4), and the energy is

$$E_0 = \int \mu \epsilon^2 (u_\phi^0, u_\phi^0) dV$$ (30)

$$\approx 10^{46} \mathrm{ergs} \left( \frac{\langle B^2 \rangle^{1/2}}{10^{15} \mathrm{G}} \right)^2 \left( \frac{R_*}{10^6 \mathrm{cm}} \right)^3 \left( \frac{a/R_*}{0.01} \right)^2 \frac{I}{10^2},$$

where $\epsilon^2$ is the contraction of the strain tensor for the Alfvén waves, $a$ is the initial condition amplitude, and $I$ is a geometric factor that is typically 10 to $10^3$. The amplitude of a mode with $\ell$ and $n$ is

$$u_{\ell n}^{\phi >} = \pi \, \mathrm{Im}(\omega_{\ell n}) \sum_{m=-\ell}^{\ell} a_{\ell m} (\mathrm{Re}(\omega_{\ell n}))$$

$$\times \quad \bar{u}_{\ell m}^{\phi >} \left( r_p, \mathrm{Re}(\omega_{\ell n}) \right).$$ (31)

Combining a mode's amplitude with its corresponding frequency as a coordinate pair,

$$\bar{u}_s \equiv \left( \mathrm{Re}(\omega_{\ell n}), u_{\ell n}^{\phi >} \right),$$ (32)

for all $\ell$ and $n$ within the desired frequency range gives the spectra shown in Figs. 5, 7, 8, and 9.

## 5 CONSTRAINTS ON GEOMETRY AND ENERGETICS

### 5.1 Timescale and amplitude considerations

The $\lesssim 4$ ms rise in luminosity observed in giant flares requires a quick release of energy within that timescale. While the time scale arguments of Lyutikov (2006) and Link (2014) indicate that a significant amount of the flare energy must be deposited directly into the magnetosphere, we can obtain further constraints by requiring that the deposited magnetic energy produces sufficient oscillation amplitude at the stellar surface to produce the modulation of the QPOs within the flare energy budget of $4 \times 10^{43}$ ergs. In the model of Timokhin et al. (2008), the required amplitude is $\gtrsim 0.01 R_*$ at the stellar surface, which we adopt as our nominal value (see §3).

We do not study the conversion of Afvén energy to radiation. Also, we need not specify exactly how the magnetic energy is converted to wave energy. We assume only that Alvén energy is deposited *suddenly*, with some degree of localization. Hence, our simulations with energy deposition in the magnetosphere could represent the tearing mode instability proposed by Lyutikov (2006) (over a time scale of $\mu s$). Likewise, our simulations with energy deposited inside the star could represent failure of the crust (over a time scale of $ms$), as in the model of Thompson & Duncan (1995) and Thompson & Duncan (2001), or some unspecified magnetic instability in the stellar core.

We find generally that for energy deposition localized entirely within the star, a surface amplitude of $\sim 0.01 R_*$ is attainable, but the energy reaches $\sim 2 R_*$ only very slowly because of the impedance mismatch discussed in §2.1. Hence, the energy is not available to create the pair fireball. On the other hand, if the energy is deposited entirely in the magnetosphere, most of the energy reflects off the star, exciting the surface motion to very small amplitude, another manifestation of the impedance mismatch. We are led to the conclusion that energy must be deposited both inside the star and in the magnetosphere so that a surface oscillation amplitude of $\sim 0.01 R_*$ can be reached without too much energy remaining trapped in the star. We now present examples of this general behavior.

Interior Origin

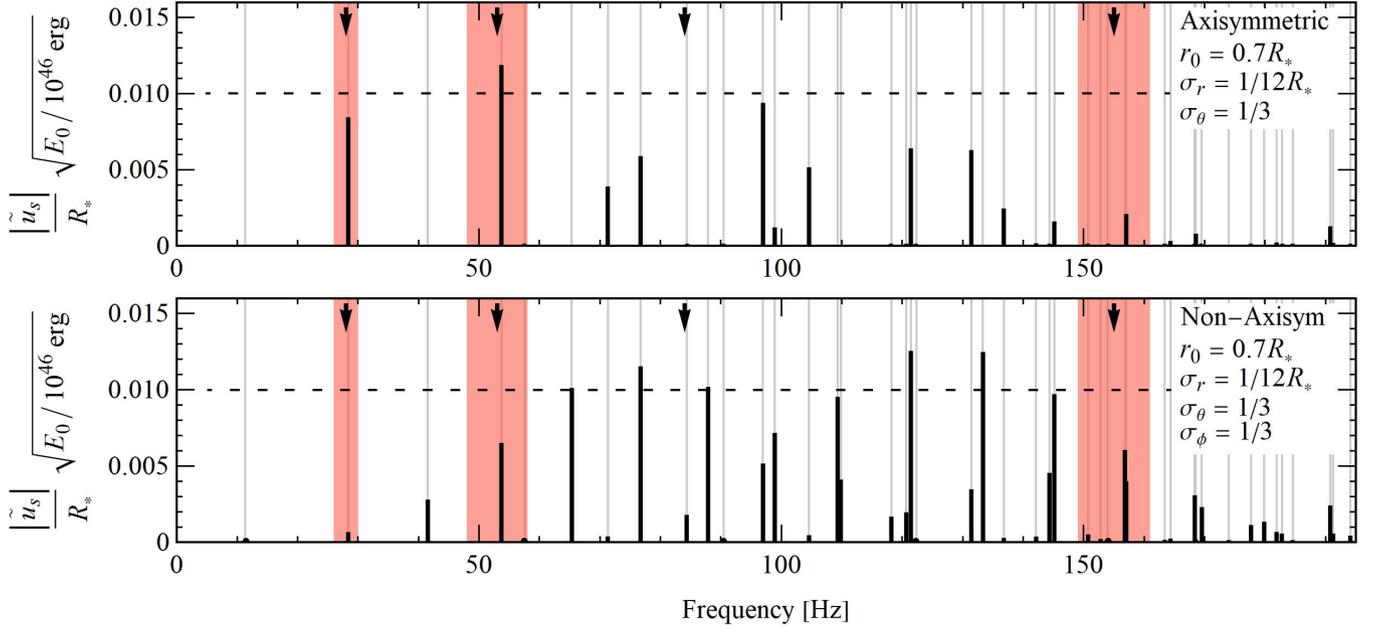

**Figure 5.** Square root of Fourier power for two initial conditions normalized to $10^{46}$ ergs initial elastic energy with energy deposited entirely inside the star. The amplitude of a peak indicates the amplitude of displacement of the magnetic field at the probe point on the stellar surface, *e.g.* $\boldsymbol{r}_p = (R_*, \theta_p, \phi_p)$. The initial conditions' parameters are $r_0 = 0.7R_*$ and $\theta_0 = \pi/2$, and for the non-axisymmetric initial condition $\phi_0 = \pi$. The black arrows indicate observed QPOs from SGR 1900's giant flare, and the pink bands show measured QPO widths. The vertical gray lines represent type-I modes: $\mathrm{Re}\left[\omega_{\ell n}\right]/2\pi$ from eq.(14). The horizontal dashed lines demarcate the observability threshold from §3. **Top:** An axisymmetric initial condition with energy deposited inside, near the surface of the star shows selective excitation of modes in comparison with the amount of modes available to the system. **Bottom:** A localized, non-axisymmetric deposition of energy also located inside, near the surface of the star excites about twice as many modes in comparison with the axisymmetric case.

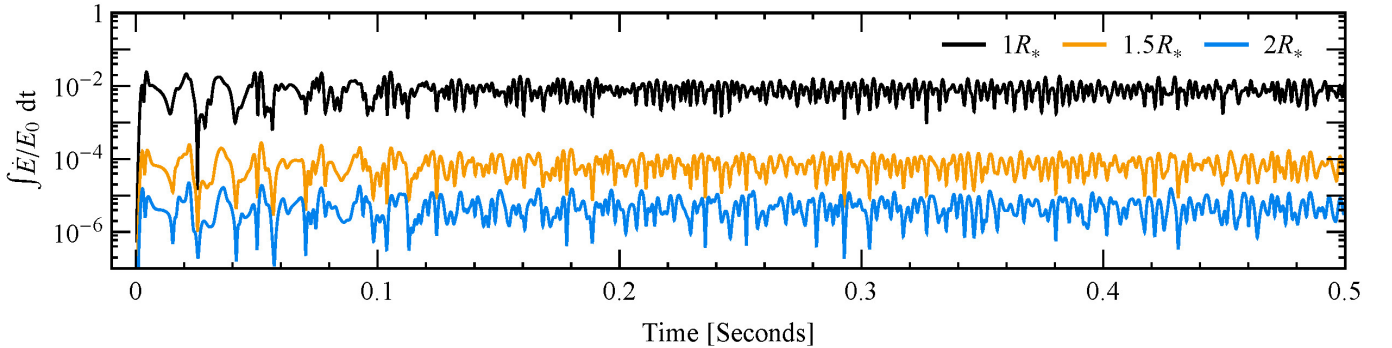

**Figure 6.** Accumulated energy, normalized by initial energy, flowing through a spherical surface of radius $1R_*$ in black, $1.5R_*$ in gold, and $2R_*$ in blue. We use an initial condition with all of the initial energy deposited inside the star: $r_0 = 0.7R_*$, $\theta_0 = \pi/2$, $\phi_0 = \pi$, $\sigma_r = 1/12R_*$, $\sigma_\theta = 1/3$, $\sigma_\phi = 1/3$. The time series is 0.5 seconds long with a resolution of $\Delta t = 0.2$ ms. Although energy leaves the star over timescales much longer than 1 second, some energy begins flowing in and out of the star within a crossing time.

## 5.2 Localized Energy Deposition Inside the Star

Fig. 5 shows the Fourier spectrum for oscillations at the stellar surface for non-axisymmetric and axisymmetric initial conditions with the energy deposited entirely inside the star. Excitation of modes to amplitudes of $\sim 0.01R_*$ requires $\sim 10^{46}$ ergs of Alfvén energy. For the axisymmetric case, three modes are excited with $\sim 0.01R_*$ amplitude near the three lowest observed QPOs. The spectrum of excited modes is more sparse for the axisymmetric case, due to the high degree of symmetry.

We attribute the 28 Hz QPO to the $\ell = 2$ fundamental, as in LvE, which implies an interior tangled field strength of $2.3 \times 10^{15}$, about three times the inferred dipole component of $7 \times 10^{14}$ G, and ten times the magnetic energy density. This identification self-consistently implies a field strength that justifies the approximation of a strong field.

Fig. 6 shows the total wave energy arriving in the magnetosphere

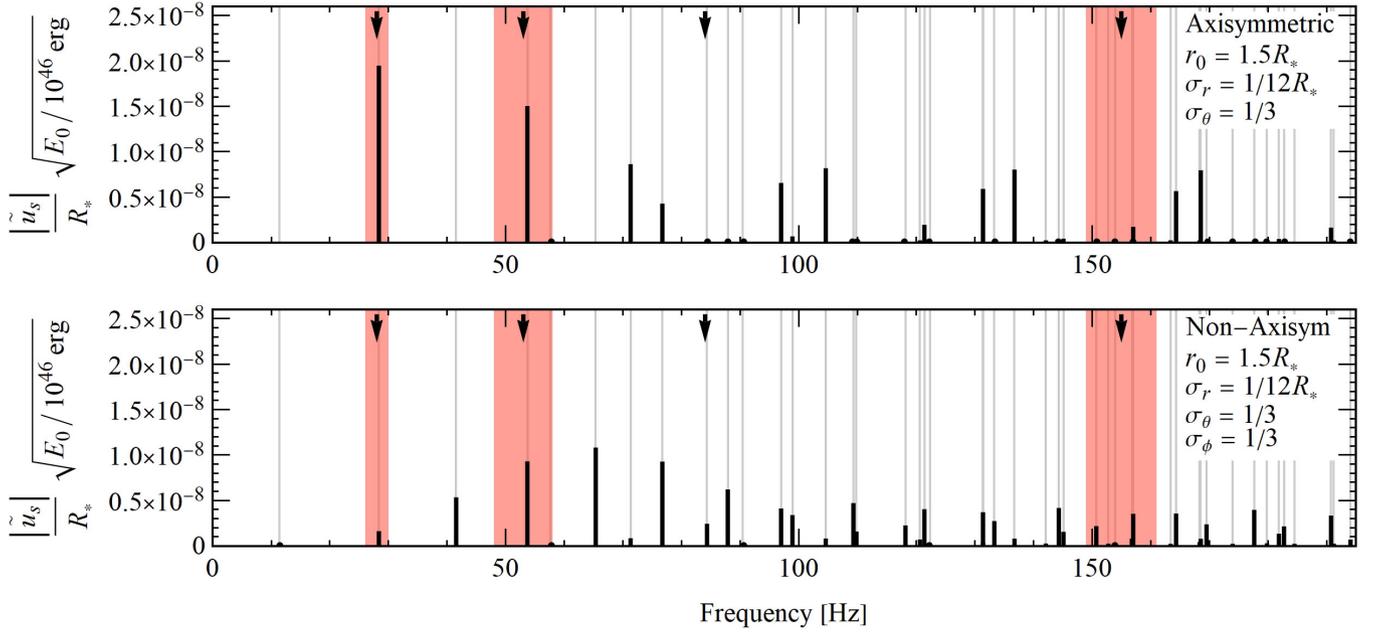

**Figure 7.** The same as Fig. 5 but the energy is deposited entirely outside the star, in the magnetosphere, radially centered at $r_0 = 1.5R_*$. The amplitude of displacement of modes for a magnetospheric flare is six orders of magnitude below the interior and combined cases.

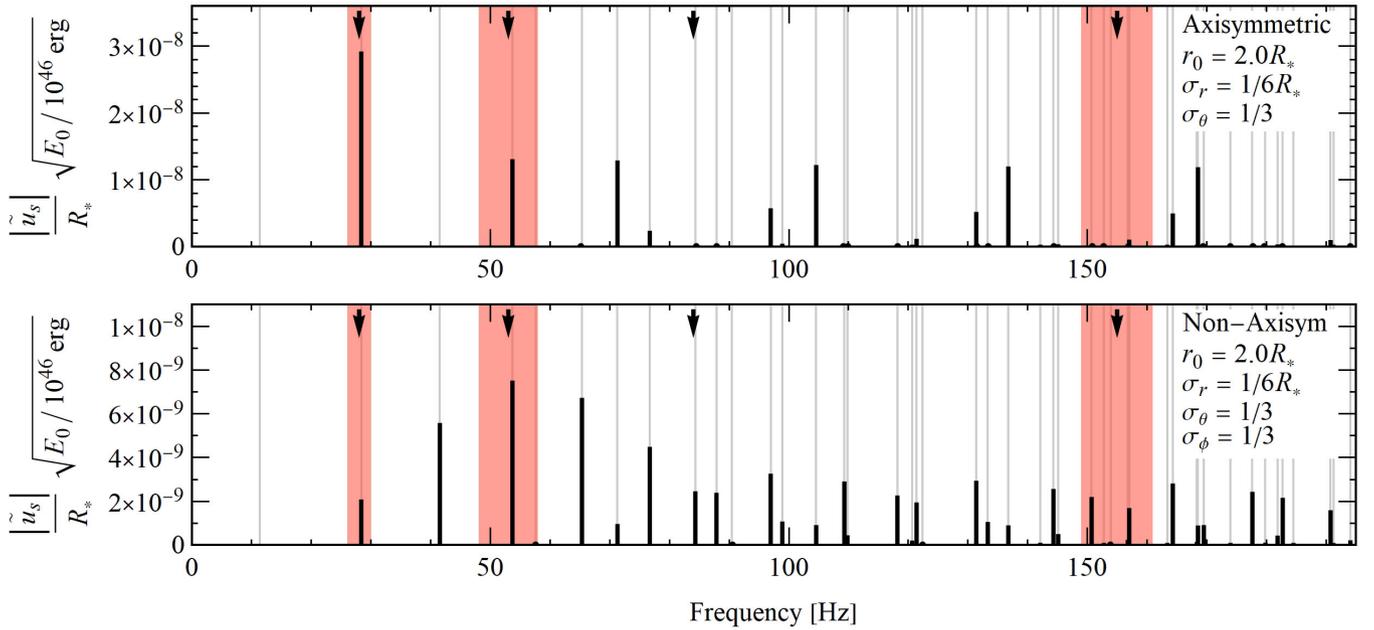

**Figure 8.** The same as Fig. 7 but the energy is deposited farther outside the star, radially centered at $r_0 = 2.0R_*$, and radially wider $\sigma_r = 1/6R_*$. The amplitude of displacement of modes is shifted slightly to lower frequencies for this case, and is also six orders of magnitude below the interior and combined cases.

Combined Interior/Magnetospheric Origin

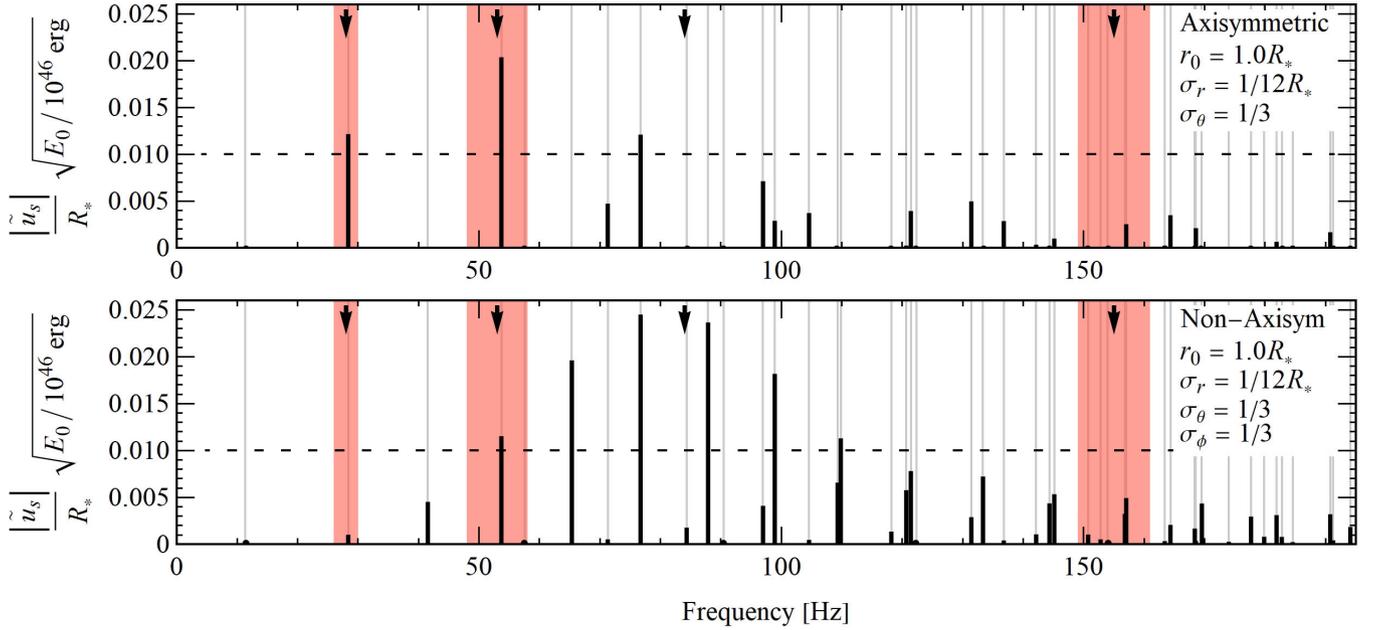

**Figure 9.** The same as Fig. 5 but the energy is deposited both inside and outside the star, straddling the stellar surface, radially centered at $r_0 = 1.0R_*$. The excitation of modes for the combined origin case is similar but higher than the amplitudes of the interior origin case. This is due to the external strain energy of the initial condition contributing less to the $10^{46}$ ergs energy budget which in turn requires larger amplitude.

during the first 0.5 s after the energy deposition, in units of the total energy $E_0$ that is deposited. The total Alfvén energy arriving in the magnetosphere is available for conversion to photons through production of a fireball. As a result of the impedance mismatch discussed in §2.1, only a fraction ($10^{-4}$) of the energy arrives at $r = 1.5R_*$ in 0.5 s, and a fraction ($10^{-6}$) at $r = 2R_*$. A small fraction of the energy exits the star over the relatively quick core crossing time of $\sim 50$ ms, but then remains trapped near the stellar surface. Though not shown on the plot, it takes years for a significant fraction of $E_0$ to propagate beyond $1.5R_*$.

The restricted flow of Alfvén energy out of the star makes energy release confined to the interior of the star difficult to reconcile with the observed quick rise in luminosity at the start of a giant flare where about half or more of the observed energy is released in the initial pulse (Mazets et al. 1999; Hurley et al. 2005). Regarding SGR 1900, an interior-origin giant flare must release at least $\sim 10^{46}$ ergs of mechanical energy that is converted in milliseconds at 100% efficiency into radiated energy to match the observed initial luminosity pulse. We conclude the flare cannot be powered by the extraction of mechanical energy from inside the star, because insufficient energy escapes the star promptly, in agreement with the conclusions of Link (2014).

### 5.3 Localized Energy Deposition in the Magnetosphere

For external energy deposition, the amplitude of the plasma is much larger, but the plasma must move the stellar surface whose mass per unit volume is larger by a factor $\rho_< / \rho_> \simeq 10^6$. The energy is in modes up to a cut-off frequency that is $\sim 300$ times larger than if the energy is deposited inside the star, and the surface amplitude is significantly reduced.

Examples of Fourier spectra of surface oscillations for a deposi-tion of energy entirely outside the star are shown in Figs. 7 and 8. Now there is no difficulty accounting for the quick rise in radiative luminosity over $\sim 4$ ms, since waves propagate across the magnetosphere in $\sim R_*/c \approx 30\mu s$, but now most of the energy reflects off the stellar surface, exciting the surface to amplitudes of only $\sim 10^{-8}R_*$ for $E_0 = 10^{46}$ ergs. To reach the nominal detectability threshold, $E_0 = 10^{57}$ ergs is required, far greater than the total magnetic energy of the star. The high reflectivity of the stellar surface to waves is another manifestation of the impedance mismatch discussed in §2.1.

We conclude that deposition of energy directly into the magnetosphere cannot excite stellar oscillations to sufficient amplitude to explain observed QPOs.

### 5.4 Localized Energy Deposition Near the Stellar Surface

Fig. 9 shows examples for energy deposition that straddles the stellar surface. The nominal detectability threshold is reached for an energy deposition of $E_0 = 10^{46}$ ergs. The portion of energy deposited outside is available within $\sim R_*/c = 30\mu s$ for creation of the fireball, while the remainder of the energy excites oscillations of the stellar surface at relatively high amplitude.

### 6 DISCUSSION AND CONCLUSIONS

Lyutikov (2006) and Link (2014) recognized that the quick rise in photon luminosity seen in giant flares requires that a significant fraction of the magnetic energy is released directly into the magnetosphere. The new feature of our analysis is evaluation of the oscillation amplitude at the stellar surface, leading us to conclude that a large fraction of the energy that drives the flare must be deposited inside the star to give sufficient amplitude to produce the observed QPOs.

Possibly the gradual unwinding of the interior magnetic field winds the field in the magnetosphere until an instability (such as the tearing mode) releases magnetic energy, while also causing the crust to fail. Thompson & Duncan (1995, 2001) attributed the $\sim 1$ s width of the peak in luminosity of SGR 1900's giant flare with the time scales required for wave energy to emerge from the stellar interior. Such a time scale does not appear in our model. This time scale could be related to the evolution of the pair fireball. Lyutikov (2006) attributes the initial spike width of $\sim 100$ ms with the dynamical time of an expanding relativistic, magnetic cloud.

For the tangled field model developed in this paper, association of the $\ell = 2$ fundamental mode with the observed 28 Hz QPO implies an interior field $\sim 2.3 \times 10^{15}$ G, about three times the inferred dipole component of $7 \times 10^{14}$ G which determines the star's spin-down rate, and ten times the magnetic energy density. This value is consistent with the simulations of Braithwaite (2009), which identified stable equilibria with a factor of 10 or more energy in the tangle than in the dipolar component. It is of course possible that the $\ell = 2$ fundamental mode has not been detected, which would imply a lower interior field.

The nominal detectability threshold for surface oscillations is reached in SGR 1900 only if $\sim 10^{46}$ ergs is deposited. This energy requirement falls within the total magnetic energy budget of $\sim 10^{48}$ (for an interior tangled field of strength $2.3 \times 10^{15}$ G). The total energy released as radiation in SGR 1900's giant flare was $\gtrsim 4 \times 10^{43}$ ergs, implying an efficiency for the conversion of wave energy to radiation (via a fireball) as low as 0.4%.

For non-axisymmetric energy deposition, we find excitation of many more modes than the four QPOs observed. Some of these modes are close together in frequency, and might be observed as a single QPO. Axisymmetric energy deposition produces a spectrum of modes that is much less dense, because of the high degree of symmetry. A possible trigger for an axisymmetric displacement is the sudden rearrangement of a twisted magnetosphere (Akgün et al. 2017, 2018; Mahlmann et al. 2019). Mahlmann et al. find a nearly axisymmetric instability develops over a few ms leading to the propagation of energy towards the star creating enough stress to potentially fracture the crust. This trigger mechanism leads to a release of substantial energy in the magnetosphere, but most of that energy will reflect off of the star and dissipate via external currents. If the crust also fails as the instability develops, then a substantial amount of energy could be released inside the star.

The large energy requirement for modes to be detectable in our model can only be met for giant flares. Less energetic events such as intermediate flares and short bursts should not be able to excite QPOs to a detectable level. QPOs have yet to be found in intermediate flares (Watts 2011), however Huppenkothen et al. (2014a) found two QPOs at $\sim$93 Hz and one at 127 Hz in averaged short recurring bursts from SGR J1550–5418. Huppenkothen et al. (2014b) also found a candidate QPO detection at 57 Hz in 30 averaged short bursts from SGR 1806–20. Short bursts are many orders of magnitude less energetic than giant flares spanning $\sim 10^{36}$ to $10^{42}$ ergs. It is possible the short-burst QPOs are from stellar oscillations that were excited by a previous giant flare. In our model, many modes persist for years with very slow damping through electron viscosity and the escape of Alfvén energy to infinity.

Miller et al. (2019) found that no QPO from the giant flare of SGR 1806–20 (an object with an implied stronger magnetic field and more energetic giant flare than SGR 1900) persists longer than $\sim 1$ second. Their results suggest excited modes are quickly damped, possibly via resonant absorption, and therefore repeatedly excited as the flare continues. Alternatively, if modes are excited at the onset of the giant flare and persist throughout the duration of the flare tail, then some mechanism must explain the observed QPO damping. A similar analysis of the SGR 1900's giant flare will shed more light on the nature of QPO damping.

Our treatment of the magnetic field of the star is rather crude. We treated the interior field as an isotropic tangle that jumps at the stellar surface to the dipole value inferred from spin down. Our treatment does not handle the transition to an ordered magnetic field as the multipoles above dipole order decay. At this point it is difficult to assess just how drastic this assumption might be, but we expect that more realistic magnetic field configurations will not substantially alter the mode spectrum we have calculated, the degree of excitation of these modes, or the Alfvén energy luminosity at infinity.

## ACKNOWLEDGMENTS

We thank D. Longcope for useful discussions, and the anonymous referee for useful criticism. This work was supported by NSF award AST-1211391 and NASA awards NNX12AF88G and NNX17AJ45G. J.B. was supported by an MSGC graduate fellowship.

## DATA AVAILABILITY

No new data were generated or analysed in support of this research.

## APPENDIX A:  ANGULAR MOMENTUM

The initial condition, see eqs.(27) and (28), is assumed to be of the form,

$$\boldsymbol{u}^0 = -\nabla \times \left( f_r f_\theta' f_\phi' \boldsymbol{r} \right) , \tag{A1}$$

where

$$f_r(r) \equiv \frac{r}{R_*} e^{-\frac{(r-\bar{r}_0)^2}{2\sigma_r^2}} \qquad \text{with} \quad \bar{r}_0 \equiv (r_0^2 - \sigma_r^2)/r_0 , \tag{A2}$$

$$f_\theta(\theta) \equiv \sin^2(\theta) e^{-\frac{(\theta-\theta_0)^2}{2\sigma_\theta^2}} , \tag{A3}$$

$$f_\phi(\phi) \equiv e^{-\frac{(\phi-\phi_0)^2}{2\sigma_\phi^2}} . \tag{A4}$$

The specific angular momentum is $\boldsymbol{j} = \rho \boldsymbol{r} \times \boldsymbol{u}^0$. Using the result $\nabla_i r_j = \delta_{ij}$ the angular momentum is

$$j_i = \rho \nabla_j \left[ f_r f_\theta' f_\phi' \left( r_i r_j - \delta_{ij} r^2 \right) \right] - 2\rho f_r f_\theta' f_\phi' r_i . \tag{A5}$$

Integrating over the volume and applying Gauss' Law yields

$$J_i = \int j_i dV = \rho \int \left[ f_r f_\theta' f_\phi' \left( r_i r_j - \delta_{ij} r^2 \right) \right] \hat{r}_i dA - 2\rho \int f_r f_\theta' f_\phi' r_i dV . \tag{A6}$$

The first term vanishes and the second must be evaluated component by component. Projecting out the $x$, $y$ and $z$ components of $\boldsymbol{r}$ yields

$$J_x = -2\rho \int f_r r^3 dr \int f_\theta' \sin^2\theta d\theta \int f_\phi' \cos\phi d\phi , \tag{A7}$$

$$J_y = -2\rho \int f_r r^3 dr \int f_\theta' \sin^2\theta d\theta \int f_\phi' \sin\phi d\phi , \tag{A8}$$

$$J_z = -2\rho \int f_r r^3 dr \int f_\theta' \sin\theta \cos\theta d\theta \int f_\phi' d\phi . \tag{A9}$$

For an initial condition to have zero angular momentum, the above integrals must all vanish. This is achieved by using symmetry arguments.

For axisymmetric initial conditions $f_\phi' =$ constant and $J_x = J_y = 0$. For rigid body rotation $f_r = r\Omega$ and $f_\theta' = -\cos\theta$ giving

$$J_z = \frac{8\pi}{15} \rho \Omega R^5 , \tag{A10}$$

which is the correct result for a rotating rigid sphere.

## APPENDIX B:  CONSERVATION OF ENERGY

The system obeys the energy conservation law in both regions,

$$\frac{\partial}{\partial t} \left( \frac{1}{2} \rho \dot{u}^2 + \mu \epsilon^2 \right) = 2\mu \nabla_i \left( \dot{u}_j \epsilon_{ij} \right) . \tag{B1}$$

where $\mu$ is the effective shear modulus of the tangled magnetic field. The effective mass density inside the star is $\rho_< = \rho_d$ from eq.(10). For relativistic Alfvén waves outside the star, $\rho_> = \mu_>/c^2$ where $\mu_> = B_d^2/4\pi$. The strain tensor is

$$\epsilon_{ij} = \frac{1}{2} \left( \nabla_i u_j + \nabla_j u_i \right) . \tag{B2}$$

In spherical coordinates the strain tensor for the Alfvén waves is

$$\epsilon_{rr} = 0 , \qquad \epsilon_{\theta\theta} = \frac{1}{r} \frac{\partial u_\theta}{\partial \theta} , \qquad \epsilon_{\phi\phi} = \frac{1}{r\sin\theta} \frac{\partial u_\phi}{\partial \phi} + \frac{u_\theta \cot\theta}{r} ,$$

$$\epsilon_{r\theta} = \frac{r}{2} \frac{\partial}{\partial r} \left( \frac{u_\theta}{r} \right) , \qquad \epsilon_{r\phi} = \frac{r}{2} \frac{\partial}{\partial r} \left( \frac{u_\phi}{r} \right) , \qquad \epsilon_{\theta\phi} = \frac{\sin\theta}{2r} \frac{\partial}{\partial \theta} \left( \frac{u_\phi}{\sin\theta} \right) + \frac{1}{2r\sin\theta} \frac{\partial u_\theta}{\partial \phi} . \tag{B3}$$

The strain energy in a volume is

$$\int \mu \epsilon^2 dV = \int \mu \left( \epsilon_{\theta\theta}^2 + \epsilon_{\phi\phi}^2 + 2\epsilon_{r\theta}^2 + 2\epsilon_{r\phi}^2 + 2\epsilon_{\theta\phi}^2 \right) dV . \tag{B4}$$

The energy flux through a surface of constant radius is

$$\dot{E} = \int 2\mu \nabla_i \left( \dot{u}_j \epsilon_{ij} \right) dV = 2\mu \int \hat{r}_i \dot{u}_j \epsilon_{ij} dS$$
$$= 2\mu \int \left( \dot{u}_\theta \epsilon_{r\theta} + \dot{u}_\phi \epsilon_{r\phi} \right) dS . \tag{B5}$$

Substituting the expressions for the strain tensor elements and velocity and applying the angular orthogonality relations with the angular integrals, eq.(C8), the energy flux through a spherical surface with radius $r > R_*$ is

$$\dot{E}_> = \mu_> \sum_{\ell=1}^{\infty} \ell(\ell+1) \sum_{m=-\ell}^{\ell} \left[ \int_0^{2\pi} \int_0^{\pi} \left( \frac{\partial^2 f_\theta}{\partial\theta^2} \frac{\partial f_\phi}{\partial\phi} \frac{\partial Y^*_{\ell m}}{\partial\theta} + \frac{1}{\sin^2\theta} \frac{\partial f_\theta}{\partial\theta} \frac{\partial^2 f_\phi}{\partial\phi^2} \frac{\partial Y^*_{\ell m}}{\partial\phi} \right) \sin\theta d\theta d\phi \right]^2 \left[ r^2 \frac{\partial F_\ell}{\partial r} - r F_\ell \right] \frac{\partial F_\ell}{\partial t} \,, \tag{B6}$$

where $F_\ell(r, t)$ is defined in Appendix D, see eq.(D2).

# APPENDIX C: ORTHOGONALITY RELATIONS

Solving for $A_\ell$ and $B_\ell$ using (6) and (7) yields

$$
\begin{aligned}
A_\ell(\omega) &= \left( \frac{iR_*\omega}{c_>} \right) \left\{ j_\ell \left( \frac{R_*\omega}{c_<} \right) \left[ \left( \frac{R_*\omega}{c_>} \right) h^{(2)}_{\ell-1} \left( \frac{R_*\omega}{c_>} \right) - (\ell+2) h^{(2)}_\ell \left( \frac{R_*\omega}{c_>} \right) \right] \right. \\
&\quad - \left. \frac{\mu_<}{\mu_>} h^{(2)}_\ell \left( \frac{R_*\omega}{c_>} \right) \left[ \left( \frac{R_*\omega}{c_<} \right) j_{\ell-1} \left( \frac{R_*\omega}{c_<} \right) - (\ell+2) j_\ell \left( \frac{R_*\omega}{c_<} \right) \right] \right\} \,,
\end{aligned}
\tag{C1}
$$

$$
\begin{aligned}
B_\ell(\omega) &= -\left( \frac{iR_*\omega}{c_>} \right) \left\{ j_\ell \left( \frac{R_*\omega}{c_<} \right) \left[ \left( \frac{R_*\omega}{c_>} \right) h^{(1)}_{\ell-1} \left( \frac{R_*\omega}{c_>} \right) - (\ell+2) h^{(1)}_\ell \left( \frac{R_*\omega}{c_>} \right) \right] \right. \\
&\quad - \left. \frac{\mu_<}{\mu_>} h^{(1)}_\ell \left( \frac{R_*\omega}{c_>} \right) \left[ \left( \frac{R_*\omega}{c_<} \right) j_{\ell-1} \left( \frac{R_*\omega}{c_<} \right) - (\ell+2) j_\ell \left( \frac{R_*\omega}{c_<} \right) \right] \right\} \,,
\end{aligned}
\tag{C2}
$$

where we have used the recursion relations for the spherical Bessel functions and the Wronskian of the spherical Hankel functions,

$$h^{(1)}_{\ell-1}(z) h^{(2)}_\ell(z) - h^{(2)}_{\ell-1}(z) h^{(1)}_\ell(z) = 2iz^{-2} \,. \tag{C3}$$

The two coefficients are related via:

$$B_\ell(\omega) = A_\ell(-\omega) \,, \qquad B_\ell(\omega) = A_\ell(\omega^*)^* \,. \tag{C4}$$

Writing the orthogonality condition in terms of the eigenfunctions, eqs.(19), we find

$$
\begin{aligned}
& \rho_< \int_{r<R_*} \bar{\mathbf{u}}_< (\mathbf{r}, \omega) \cdot \bar{\mathbf{u}}^*_< (\mathbf{r}, \omega') \, dV + \rho_> \int_{r>R_*} \bar{\mathbf{u}}_> (\mathbf{r}, \omega) \cdot \bar{\mathbf{u}}^*_> (\mathbf{r}, \omega') \, dV \\
&= \left[ \rho_< \int w^<_\ell \left( \frac{r\omega}{c_<} \right) w^<_{\ell'} \left( \frac{r\omega'}{c_<} \right) r^2 dr + \rho_> \int w^>_\ell \left( \frac{r\omega}{c_>} \right) w^>_{\ell'} \left( \frac{r\omega'}{c_>} \right) r^2 dr \right] \\
&\quad \times \int_0^{2\pi} \int_0^{\pi} \left[ \left( \frac{\partial Y_{\ell m}}{\partial\theta} \right) \left( \frac{\partial Y^*_{\ell' m'}}{\partial\theta} \right) + \frac{1}{\sin^2\theta} \left( \frac{\partial Y_{\ell m}}{\partial\phi} \right) \left( \frac{\partial Y^*_{\ell' m'}}{\partial\phi} \right) \right] \sin\theta d\theta d\phi \,.
\end{aligned}
\tag{C5}
$$

The $Y_{\ell m}$ satisfy the differential equation

$$\frac{1}{\sin\theta} \frac{\partial}{\partial\theta} \left[ \sin\theta \frac{\partial Y_{\ell m}}{\partial\theta} \right] + \frac{1}{\sin^2\theta} \frac{\partial^2 Y_{\ell m}}{\partial\phi^2} + \ell(\ell+1) Y_{\ell m} = 0 \,, \tag{C6}$$

with normalization

$$\int_0^{2\pi} \int_0^{\pi} Y_{\ell m} Y^*_{\ell' m'} \sin\theta d\theta d\phi = \delta_{\ell\ell'} \delta_{mm'} \,. \tag{C7}$$

Combining eqs.(C6) and (C7), integrating by parts and noting the vanishing boundary terms we obtain

$$\int_0^{2\pi} \int_0^{\pi} \left[ \left( \frac{\partial Y_{\ell m}}{\partial\theta} \right) \left( \frac{\partial Y^*_{\ell' m'}}{\partial\theta} \right) + \frac{1}{\sin^2\theta} \left( \frac{\partial Y_{\ell m}}{\partial\theta} \right) \left( \frac{\partial Y^*_{\ell' m'}}{\partial\theta} \right) \right] \sin\theta d\theta d\phi = \ell(\ell+1) \delta_{\ell\ell'} \delta_{mm'} \,. \tag{C8}$$

We now obtain an orthogonality relation for the radial dependence. We form the orthogonality condition and integrate by parts the spherical Bessel equation, eq.(5), to get

$$
\begin{aligned}
&\left( \omega^2 - \omega'^2 \right) \left[ \rho_< \int_0^{R_*} w^<_\ell(r, \omega) w^<_\ell(r, \omega') r^2 dr + \rho_> \int_{R_*}^R w^>_\ell(r, \omega) w^>_\ell(r, \omega') r^2 dr \right] \\
&= R^2 w^>_\ell(R, \omega) \mu_> \frac{\partial w^>_\ell}{\partial r}(R, \omega') - R^2 w^>_\ell(R, \omega') \mu_> \frac{\partial w^>_\ell}{\partial r}(R, \omega) \,,
\end{aligned}
\tag{C9}
$$

where $R$ is the radius of the outer boundary to the magnetosphere. Substituting eq.(8), we have

$$
\rho_< \int_0^{R_*} w_\ell^<(r,\omega) w_\ell^<(r,\omega') r^2 dr + \rho_> \int_{R_*}^R w_\ell^>(r,\omega) w_\ell^>(r,\omega') r^2 dr
$$

$$
= \frac{\mu_> R^2}{4c_> (\omega^2 - \omega'^2)} \left\{ A_\ell(\omega') A_\ell(\omega) \left[ \omega h_{\ell+1}^{(1)}\left(\frac{R\omega'}{c_>}\right) h_\ell^{(1)}\left(\frac{R\omega}{c_>}\right) - \omega' h_\ell^{(1)}\left(\frac{R\omega}{c_>}\right) h_{\ell+1}^{(1)}\left(\frac{R\omega'}{c_>}\right) \right] \right.
$$

$$
+ \; B_\ell(\omega') A_\ell(\omega) \left[ \omega h_\ell^{(2)}\left(\frac{R\omega'}{c_>}\right) h_{\ell+1}^{(1)}\left(\frac{R\omega}{c_>}\right) - \omega' h_\ell^{(1)}\left(\frac{R\omega}{c_>}\right) h_{\ell+1}^{(2)}\left(\frac{R\omega'}{c_>}\right) \right]
$$

$$
+ \; A_\ell(\omega') B_\ell(\omega) \left[ \omega h_\ell^{(1)}\left(\frac{R\omega'}{c_>}\right) h_{\ell+1}^{(2)}\left(\frac{R\omega}{c_>}\right) - \omega' h_\ell^{(2)}\left(\frac{R\omega}{c_>}\right) h_{\ell+1}^{(1)}\left(\frac{R\omega'}{c_>}\right) \right]
$$

$$
+ \; B_\ell(\omega') B_\ell(\omega) \left[ \omega h_\ell^{(2)}\left(\frac{R\omega'}{c_>}\right) h_{\ell+1}^{(2)}\left(\frac{R\omega}{c_>}\right) - \omega' h_\ell^{(2)}\left(\frac{R\omega}{c_>}\right) h_{\ell+1}^{(2)}\left(\frac{R\omega'}{c_>}\right) \right] \right\} . \tag{C10}
$$

We are interested in the limit $R \to \infty$, for which the spherical Hankel functions become

$$
h_\ell^{(1)}(z) \sim \frac{1}{z} e^{i\left[z - \frac{\pi}{2}(\ell+1)\right]} , \qquad h_\ell^{(2)}(z) \sim \frac{1}{z} e^{-i\left[z - \frac{\pi}{2}(\ell+1)\right]} , \tag{C11}
$$

and we obtain

$$
\rho_< \int_0^{R_*} w_\ell^<(r,\omega) w_\ell^<(r,\omega') r^2 dr + \rho_> \int_{R_*}^R w_\ell^>(r,\omega) w_\ell^>(r,\omega') r^2 dr
$$

$$
= \frac{\mu_> c_>}{4\omega\omega'} \left\{ \sin\left(\frac{R(\omega-\omega')}{c_>}\right) \left[ \frac{A_\ell(\omega') B_\ell(\omega) + B_\ell(\omega') A_\ell(\omega)}{\omega - \omega'} \right] \right.
$$

$$
+ \; i \cos\left(\frac{R(\omega-\omega')}{c_>}\right) \left[ \frac{A_\ell(\omega') B_\ell(\omega) - B_\ell(\omega') A_\ell(\omega)}{\omega - \omega'} \right]
$$

$$
+ \; (-1)^{\ell+1} \sin\left(\frac{R(\omega+\omega')}{c_>}\right) \left[ \frac{B_\ell(\omega') B_\ell(\omega) + A_\ell(\omega') A_\ell(\omega)}{\omega + \omega'} \right]
$$

$$
+ \; i(-1)^{\ell+1} \cos\left(\frac{R(\omega+\omega')}{c_>}\right) \left[ \frac{B_\ell(\omega') B_\ell(\omega) - A_\ell(\omega') A_\ell(\omega)}{\omega + \omega'} \right] \right\} . \tag{C12}
$$

Regarding the cosine terms, we expand $A_\ell(\omega') B_\ell(\omega) - B_\ell(\omega') A_\ell(\omega)$ in $\omega$ about $\omega'$ and find the leading order term is linear in $\omega - \omega'$, thus canceling the divergence at $\omega = \omega'$. The same is found for the second cosine term under the transformation $\omega' \to -\omega'$, again canceling the divergence at $\omega = -\omega'$. What's left are cosines modulated by a bound function. In the limit of $R \to \infty$, the infinitesimal oscillations of the cosines give zero when integrated against any Schwartz function. Thus the cosine terms can be dropped from the expression.

Regarding the sine terms we use the identity

$$
\delta(k) = \lim_{D\to\infty} \frac{1}{2\pi} \int_{-D}^D e^{ikx} dx = \lim_{D\to\infty} \frac{\sin(Dk)}{\pi k} , \tag{C13}
$$

to obtain

$$
\rho_< \int_0^{R_*} w_\ell^<(r,\omega) w_\ell^<(r,\omega') r^2 dr + \rho_> \int_{R_*}^\infty w_\ell^>(r,\omega) w_\ell^>(r,\omega') r^2 dr
$$

$$
= \frac{\pi c_> \mu_>}{4\omega\omega'} \left\{ \left[ A_\ell(\omega') B_\ell(\omega) + B_\ell(\omega') A_\ell(\omega) \right] \delta(\omega - \omega') - (-1)^\ell \left[ A_\ell(\omega') A_\ell(\omega) + B_\ell(\omega') B_\ell(\omega) \right] \delta(\omega + \omega') \right\} . \tag{C14}
$$

If the domain only extends to $\omega > 0$, then eq.(C14) reduces to

$$
\rho_< \int_0^{R_*} w_\ell^<(r,\omega) w_\ell^<(r,\omega') r^2 dr + \rho_> \int_{R_*}^\infty w_\ell^>(r,\omega) w_\ell^>(r,\omega') r^2 dr = \frac{\pi c_> \mu_>}{2\omega^2} A_\ell(\omega) B_\ell(\omega) \delta(\omega - \omega') . \tag{C15}
$$

In the limit that $\rho_> = \rho_<$ and $\mu_> = \mu_<$ we have $c_> = c_<$, $A(\omega) = B(\omega) = 1$ and therefore $w_\ell^< = w_\ell^>$ and eq.(C15) reduces to the familiar definition of the delta function in terms of spherical Bessel functions, viz.

$$
\int_0^\infty j_\ell\left(\frac{\omega r}{c_<}\right) j_\ell\left(\frac{\omega' r}{c_<}\right) r^2 dr = \frac{\pi c_<^2}{2\omega^2} \delta\left(\frac{\omega}{c_<} - \frac{\omega'}{c_<}\right) . \tag{C16}
$$

Combining eqs.(C8) and (C15), and noting we integrate over $\omega > 0$, the required orthogonality condition is

$$
\rho_< \int_{r<R_*} \boldsymbol{u}_<(r,\omega) \cdot \boldsymbol{u}_<^*(r,\omega') \, dV + \rho_> \int_{r>R_*} \boldsymbol{u}_>(r,\omega) \cdot \boldsymbol{u}_>^*(r,\omega') \, dV
$$

$$
= \left[ \ell(\ell+1) \frac{\pi c_> \mu_>}{2\omega^2} A_\ell(\omega) B_\ell(\omega) \right] \delta_{\ell\ell'} \delta_{mm'} \delta(\omega - \omega') . \tag{C17}
$$

## APPENDIX D:  CONTOUR INTEGRALS

The spatial integrals in eq.(24) and frequency integrals in eqs.(26) are numerically calculated. In the limit of a weakly coupled magnetosphere ($\gamma \ll 1$), the integrands in eqs.(26) become computationally cumbersome. Dozens of peaks in the frequency domain become very sharp with widths approximately the imaginary component of eq.(14). To cut the computation time down by more than two orders of magnitude, we extend to the complex frequency domain and recast the frequency integrals in eqs.(26) as contour integrals. Utilizing the Residue Theorem, the solutions can be rewritten as a sum over poles in the complex frequency plane.

We proceed solving only for the magnetospheric solutions in eqs.(26). The contour integral approach to solving the interior solutions leads to a very slowly converging solution near the origin and undefined at the origin. This results from effectively changing the interior basis functions from spherical Bessel functions to spherical Hankel functions which diverge at the origin for complex argument. For this reason we exclude the contour integrals for the interior solutions which are not needed for the presented calculations.

To begin, we pull out the frequency dependence in $a_{\ell m}$, eq.(24); plug in the radial solutions, eq.(8); and rewrite $j_\ell(z) = \frac{1}{2}\left(h_\ell^{(1)}(z) + h_\ell^{(2)}(z)\right)$ to get

$$
\begin{aligned}
\bar{a}_\ell(\omega) \equiv {} & \rho_< \int_0^{R_*} f_r(r')\frac{1}{2}\left[h_\ell^{(1)}\left(\frac{\omega r'}{c_<}\right) + h_\ell^{(2)}\left(\frac{\omega r'}{c_<}\right)\right] r'^2 dr' \\
& + \rho_> \int_{R_*}^\infty f_r(r')\frac{1}{2}\left[A_\ell(\omega)h_\ell^{(1)}\left(\frac{\omega r'}{c_>}\right) + B_\ell(\omega)h_\ell^{(2)}\left(\frac{\omega r'}{c_>}\right)\right] r'^2 dr' \,.
\end{aligned} \tag{D1}
$$

The frequency integrals of the full solutions, eqs.(26), separate from the angular dependence. We define a function that isolates the frequency dependence for each $\ell$,

$$
F_\ell(r,t) \equiv \frac{1}{\pi c_> \mu_> \ell(\ell+1)} \int_0^\infty \frac{\omega^2 \bar{a}_\ell(\omega)}{B_\ell(\omega)A_\ell(\omega)} \left[A_\ell(\omega)h_\ell^{(1)}\left(\frac{\omega r}{c_>}\right) + B_\ell(\omega)h_\ell^{(2)}\left(\frac{\omega r}{c_>}\right)\right]\cos(\omega t)\,d\omega\,. \tag{D2}
$$

To close the contours for integrating, we need the integrand to vanish at positive complex infinity, $i\omega \to \infty$. Thus $\bar{a}_\ell(\omega)$ must be included in the full expression,

$$
\begin{aligned}
F_\ell(r,t) = {} & \frac{\rho_<}{2\pi c_> \mu_> \ell(\ell+1)} \int_0^\infty \int_0^{R_*} f_r(r')r'^2 \left[h_\ell^{(1)}\left(\frac{\omega r'}{c_<}\right) + h_\ell^{(2)}\left(\frac{\omega r'}{c_<}\right)\right] dr' \\
& \times \frac{\omega^2}{B_\ell(\omega)A_\ell(\omega)} \left[A_\ell(\omega)h_\ell^{(1)}\left(\frac{\omega r}{c_>}\right) + B_\ell(\omega)h_\ell^{(2)}\left(\frac{\omega r}{c_>}\right)\right]\cos(\omega t)\,d\omega \\
& + \frac{\rho_>}{2\pi c_> \mu_> \ell(\ell+1)} \int_0^\infty \int_0^\infty f_r(r')r'^2 \left[A_\ell(\omega)h_\ell^{(1)}\left(\frac{\omega r'}{c_<}\right) + B_\ell(\omega)h_\ell^{(2)}\left(\frac{\omega r'}{c_<}\right)\right] dr' \\
& \times \frac{\omega^2}{B_\ell(\omega)A_\ell(\omega)} \left[A_\ell(\omega)h_\ell^{(1)}\left(\frac{\omega r}{c_>}\right) + B_\ell(\omega)h_\ell^{(2)}\left(\frac{\omega r}{c_>}\right)\right]\cos(\omega t)\,d\omega\,. 
\end{aligned} \tag{D3}
$$

Performing the change of variable $\omega \to -\omega$ on the $h_\ell^{(1)}(\omega r')$ terms, using the symmetry properties $h_\ell^{(1)}(-\omega) = (-1)^\ell \, h_\ell^{(2)}(\omega)$, $A_\ell(\omega) = B_\ell(-\omega)$, and recombining to form integrals across the entire real-frequency axis yields

$$
\begin{aligned}
F_\ell(r,t) \equiv {} & \frac{\rho_<}{4\pi c_> \mu_> \ell(\ell+1)} \int_{-\infty}^\infty \int_0^{R_*} f_r(r')r'^2 \left[h_\ell^{(1)}\left(\frac{\omega r'}{c_<}\right)\right] dr' \\
& \times \frac{\omega^2}{B_\ell(\omega)A_\ell(\omega)} \left[A_\ell(\omega)h_\ell^{(1)}\left(\frac{\omega r}{c_>}\right) + B_\ell(\omega)h_\ell^{(2)}\left(\frac{\omega r}{c_>}\right)\right]\left(e^{i\omega t} + e^{-i\omega t}\right) d\omega \\
& + \frac{\rho_>}{4\pi c_> \mu_> \ell(\ell+1)} \int_{-\infty}^\infty \int_{R_*}^\infty f_r(r')r'^2 \left[A_\ell(\omega)h_\ell^{(1)}\left(\frac{\omega r'}{c_>}\right)\right] dr' \\
& \times \frac{\omega^2}{B_\ell(\omega)A_\ell(\omega)} \left[A_\ell(\omega)h_\ell^{(1)}\left(\frac{\omega r}{c_>}\right) + B_\ell(\omega)h_\ell^{(2)}\left(\frac{\omega r}{c_>}\right)\right]\left(e^{i\omega t} + e^{-i\omega t}\right) d\omega\,. 
\end{aligned} \tag{D4}
$$

The zeros in the denominator are simple poles, and we can integrate using the residue theorem. The zeros of $A_\ell$ lie in the upper-half plane; the zeros of $B_\ell$ are opposite the zeros of $A_\ell$ and lie in the lower-half plane.

We first consider the term

$$
\int_{-\infty}^\infty \frac{\omega^2\left[A_\ell^{(1)}(\omega) + A_\ell^{(2)}(\omega)\right]}{B_\ell(\omega)A_\ell(\omega)} h_\ell^{(1)}\left(\frac{\omega r'}{c_<}\right) h_\ell^{(1)}\left(\frac{\omega r}{c_>}\right) e^{i\omega t} d\omega\,, \tag{D5}
$$

where $A_\ell = A_\ell^{(1)} + A_\ell^{(2)}$ is separated according to its asymptotics,

$$
\begin{aligned}
A_\ell^{(1)}(\omega) &= \frac{iR_*\omega}{2c_>}\left\{ h_\ell^{(2)}\left(\frac{R_*\omega}{c_<}\right)\left[\frac{R_*\omega}{c_>}h_{\ell-1}^{(2)}\left(\frac{R_*\omega}{c_>}\right) - (2+\ell)\,h_\ell^{(2)}\left(\frac{R_*\omega}{c_>}\right)\right]\right. \\
&\quad \left. - \frac{\mu_<}{\mu_>}h_\ell^{(2)}\left(\frac{R_*\omega}{c_>}\right)\left[\frac{R_*\omega}{c_>}h_{\ell-1}^{(2)}\left(\frac{R_*\omega}{c_<}\right) - \frac{c_<}{c_>}(2+\ell)\,h_\ell^{(2)}\left(\frac{R_*\omega}{c_<}\right)\right]\right\}, \\
A_\ell^{(2)}(\omega) &= \frac{iR_*\omega}{2c_>}\left\{ h_\ell^{(1)}\left(\frac{R_*\omega}{c_<}\right)\left[\frac{R_*\omega}{c_>}h_{\ell-1}^{(2)}\left(\frac{R_*\omega}{c_>}\right) - (2+\ell)\,h_\ell^{(2)}\left(\frac{R_*\omega}{c_>}\right)\right]\right. \\
&\quad \left. - \frac{\mu_<}{\mu_>}h_\ell^{(2)}\left(\frac{R_*\omega}{c_>}\right)\left[\frac{R_*\omega}{c_>}h_{\ell-1}^{(1)}\left(\frac{R_*\omega}{c_<}\right) - \frac{c_<}{c_>}(2+\ell)\,h_\ell^{(1)}\left(\frac{R_*\omega}{c_<}\right)\right]\right\}.
\end{aligned}
\tag{D6}
$$

Using the spherical Hankel functions' asymptotic form, eqs.(C11), on eqs.(D6) gives

$$
A_\ell^{(1)}(\omega) \sim e^{-i\omega\left(\frac{R_*}{c_<}+\frac{R_*}{c_>}\right)}, \qquad A_\ell^{(2)}(\omega) \sim e^{i\omega\left(\frac{R_*}{c_<}-\frac{R_*}{c_>}\right)}.
\tag{D7}
$$

Consider the first part of the term from eq.(D5),

$$
\int_{-\infty}^{\infty}\frac{\omega^2 A_\ell^{(1)}(\omega)}{B_\ell(\omega)A_\ell(\omega)}h_\ell^{(1)}\left(\frac{\omega r'}{c_<}\right)h_\ell^{(1)}\left(\frac{\omega r}{c_>}\right)e^{i\omega t}d\omega \sim \int_{-\infty}^{\infty}e^{i\omega\left(\frac{r'}{c_<}+\frac{r}{c_>}-\frac{R_*}{c_<}-\frac{R_*}{c_>}+t\right)}d\omega.
\tag{D8}
$$

For $r'/c_< + r/c_> - R_*/c_< - R_*/c_> + t > 0$, the integrand vanishes as $\omega \to i\infty$, hence the contour can be closed in the upper-half plane. The residue theorem yields

$$
\int_{-\infty}^{\infty}\frac{\omega^2 A_\ell^{(1)}(\omega)}{B_\ell(\omega)A_\ell(\omega)}h_\ell^{(1)}\left(\frac{\omega r'}{c_>}\right)h_\ell^{(1)}\left(\frac{\omega r}{c_>}\right)e^{i\omega t}d\omega = 2\pi i\sum_n\frac{\omega_{\ell n}^2 A_\ell^{(1)}(\omega_{\ell n})}{B_\ell(\omega_{\ell n})A_\ell'(\omega_{\ell n})}h_\ell^{(1)}\left(\frac{\omega_{\ell n} r'}{c_>}\right)h_\ell^{(1)}\left(\frac{\omega_{\ell n} r}{c_>}\right)e^{i\omega_{\ell n} t},
\tag{D9}
$$

where $A_\ell' \equiv \partial A_\ell(\omega)/\partial\omega$.

For $r'/c_< + r/c_> - R_*/c_< - R_*/c_> + t < 0$, the integrand vanishes as $\omega \to -i\infty$. To maintain a sum over the same poles in the upper-half plane, we replace $\omega \to -\omega$ and close the contour in the upper-half plane as before to yield

$$
\begin{aligned}
&\int_{-\infty}^{\infty}\frac{(-\omega)^2 A_\ell^{(1)}(-\omega)}{B_\ell(-\omega)A_\ell(-\omega)}h_\ell^{(1)}\left(\frac{-\omega r'}{c_>}\right)h_\ell^{(1)}\left(\frac{-\omega r}{c_>}\right)e^{-i\omega t}(-d\omega) \\
&= \int_{-\infty}^{\infty}\frac{\omega^2 B_\ell^{(2)}(\omega)}{A_\ell(\omega)B_\ell(\omega)}h_\ell^{(2)}\left(\frac{\omega r'}{c_>}\right)h_\ell^{(2)}\left(\frac{\omega r}{c_>}\right)e^{-i\omega t}d\omega \\
&= 2\pi i\sum_n\frac{\omega_{\ell n}^2 B_\ell^{(2)}(\omega_{\ell n})}{B_\ell(\omega_{\ell n})A_\ell'(\omega_{\ell n})}h_\ell^{(2)}\left(\frac{\omega_{\ell n} r'}{c_>}\right)h_\ell^{(2)}\left(\frac{\omega_{\ell n} r}{c_>}\right)e^{-i\omega_{\ell n} t},
\end{aligned}
\tag{D10}
$$

where we define $B_\ell^{(2)}(\omega) = A_\ell^{(1)}(-\omega)$. Both eqs.(D9) and (D10) give the solution to eq.(D8) depending on the condition $r'/c_< + r/c_> - R_*/c_< - R_*/c_> + t \lessgtr 0$. The solution for the term identical to eq.(D8) but with $e^{-i\omega t}$ time dependence can be evaluated by substituting $t \to -t$ in eqs.(D9) and (D10), and the condition becomes $r'/c_< + r/c_> - R_*/c_< - R_*/c_> - t \lessgtr 0$.

The second part of eq.(D5) is

$$
\int_{-\infty}^{\infty}\frac{\omega^2 A_\ell^{(2)}(\omega)}{B_\ell(\omega)A_\ell(\omega)}h_\ell^{(1)}\left(\frac{\omega r'}{c_<}\right)h_\ell^{(1)}\left(\frac{\omega r}{c_>}\right)e^{i\omega t}d\omega \sim \int_{-\infty}^{\infty}e^{i\omega\left(\frac{r'}{c_<}+\frac{r}{c_>}+\frac{R_*}{c_<}-\frac{R_*}{c_>}+t\right)}d\omega.
\tag{D11}
$$

This is evaluated in the same way as eq.(D8), considering the cases $r'/c_< + r/c_> + R_*/c_< - R_*/c_> + t \lessgtr 0$ and defining $B_\ell^{(1)}(\omega) = A_\ell^{(2)}(-\omega)$ where $B_\ell = B_\ell^{(1)} + B_\ell^{(2)}$. Again the $e^{-i\omega t}$ term can be found by substituting $t \to -t$.

We consider the next term,

$$
\int_{-\infty}^{\infty}\frac{\omega^2\left[B_\ell^{(1)}(\omega)+B_\ell^{(2)}(\omega)\right]}{B_\ell(\omega)A_\ell(\omega)}h_\ell^{(1)}\left(\frac{\omega r'}{c_<}\right)h_\ell^{(2)}\left(\frac{\omega r}{c_>}\right)e^{i\omega t}d\omega \sim \int_{-\infty}^{\infty}e^{i\omega\left(\frac{r'}{c_<}-\frac{r}{c_>}-\frac{R_*}{c_<}+\frac{R_*}{c_>}+t\right)}d\omega + \int_{-\infty}^{\infty}e^{i\omega\left(\frac{r'}{c_<}-\frac{r}{c_>}+\frac{R_*}{c_<}+\frac{R_*}{c_>}+t\right)}d\omega.
\tag{D12}
$$

The solution to this term is found in the same way as eq.(D5), but the asymptotic form of $B_\ell$ and $h_\ell^{(2)}$ are now used.

Proceeding with this method leads to $2^3(2+3) = 40$ terms, each with a different condition that depends on $r'$, $r$, and $t$. The $2^3$ factor is the permutations of $h_\ell^{(1,2)}(r')$, $h_\ell^{(1,2)}(r)$ and $e^{\pm i\omega t}$.

Accounting for all 40 conditions can be computationally cumbersome to implement for general $r'$, $r$, $t$. For time series that are calculated at late enough times, all of the conditions with positive t-dependence are selected, and the full solution simplifies to

$$
F_\ell(r,t) = \frac{i}{c_>\mu_>\ell(\ell+1)}\sum_n\frac{\omega_{\ell n}^2 \bar{a}_\ell(\omega_{\ell n})}{A_\ell'(\omega_{\ell n})}h_\ell^{(2)}\left(\frac{\omega_{\ell n} r}{c_>}\right)e^{i\omega_{\ell n} t}.
\tag{D13}
$$

This simplifying choice corresponds to choosing a time when the star has been excited and is in a ringdown phase. The $r'$ dependence in the conditions is effectively capped by the location of significant amplitude in the initial condition, even though the $r'$ integral goes to infinity; using an initial condition with a pulse that starts farther from the star requires more time for the pulse to reach the star and the star to enter a ringdown phase.

Eq.(D13) can also be used to calculate the Alfvén luminosity, eq.(B6), at $r = R_*$ and $t = 0$ for an interior deposition, because the star is already in a ringdown phase when all of the energy is deposited inside the star. At larger radii, eq.(D13) can be used when $t > r/c_> \approx 70\,\mu s$ for $r = 2R_*$. To handle the derivatives on $F_\ell(r, t)$ in eq.(B6), we evaluate the derivatives inside the integrals before numerically solving the frequency integral.

This paper has been typeset from a TeX/LaTeX file prepared by the author.